\documentclass[preprint2,tigten,letteredappendix,appendixfloats]{aastex7}

\usepackage{hyperref}
\usepackage{romannum}
\usepackage{amsmath}
\usepackage{amsmath}
\usepackage{float}

\begin{document}

\pagenumbering{arabic}

\title{A Mysteriously Tight H$\alpha$-[O~\Romannum{3}] Correlation and Non-Case B Balmer Decrements Revealed by the Spectra from the James Webb Space Telescope NIRSpec Instrument}

\author[0000-0001-7957-6202]{Bangzheng Sun} 
\affiliation{Department of Physics and Astronomy, University of Missouri - Columbia \\
701 S College Avenue \\
Columbia, MO 65201, USA}
\email{bangzheng.sun@mail.missouri.edu}

\author[0000-0001-7592-7714]{Haojing Yan}
\affiliation{Department of Physics and Astronomy, University of Missouri - Columbia \\
701 S College Avenue \\
Columbia, MO 65201, USA}
\email{yanha@missouri.edu}

\begin{abstract}

   We report an extremely tight, linear relation between ${\rm H\alpha}$ and 
[O~\Romannum{3}] line fluxes in logarithm, discovered using a large sample of 
low and mid-resolution spectra (totaling 563) obtained by the James Webb Space 
Telescope (JWST) NIRSpec instrument in three widely separated extragalactic 
fields. While a certain correlation between ${\rm H\alpha}$ and [O~\Romannum{3}] 
is expected for star forming galaxies, such a log-linear and tight (dispersion 
of $\sim$0.1 dex) trend is hard to explain because dust reddening would 
skew any intrinsic relation between the two. Furthermore, another surprising
finding emerges from investigating the dust reddening properties of these 
galaxies. We find that the classic method of using the Balmer decrements under 
the standard Case B assumption does not work: a high fraction 
(${\sim30\%}$) of 
our objects have ${\rm H\alpha}$/${\rm H\beta}$ line ratios even smaller than 
the canonical Case B ratio of 2.86. Such a high fraction of non-Case B Balmer 
decrements is also present in other JWST and ground-based spectroscopic studies,
but the universal applicability of the Case B assumption was not questioned 
until recently. The mysterious ${\rm H\alpha}$--[O~\Romannum{3}] correlation and 
the high fraction of non-Case B Balmer decrements, which may or may not be 
related, should be further investigated to put our spectral analysis onto a more 
solid footing.

\end{abstract}

\keywords{Emission-line galaxies}

\section{Introduction} \label{sec:intro}

   Emission line diagnostics is a fundamental tool in studying the physical 
properties of star-forming galaxies, which can provide deep insights into their
star formation rates (SFRs), chemical compositions, evolutionary histories, etc.
\citep[e.g., ][]{Kewley2019}. Two emission lines, ${\rm H\alpha}$$\lambda$6563 
and [O~\Romannum{3}]$\lambda\lambda4959,5007$ doublet, are the most studied ones
because they are usually the strongest lines among all. The ${\rm H\alpha}$ line 
is widely adopted as a robust tracer of instantaneous SFR because its intensity 
is directly related to the ionizing radiation from young, high-mass stars 
\citep[][]{Kennicutt1998}. The [O~\Romannum{3}] doublet, on the other hand, is 
often used as a probe of the ISM ionization state 
\citep[e.g., ][]{Baldwin1981,KD02,Maiolino2008}. 

   The utilization of these two lines in large samples used to be confined to 
$z<4$ due to the limitation of the instrumental capabilities. As an example, the 
Keck MOSDEF spectrograph cuts off at 2.4~$\mu{\rm m}$, which can only detect 
${\rm H\alpha}$ and [O~\Romannum{3}] up to $z\approx 2.7$ and $z\approx 3.8$, 
respectively. The advent of the James Webb Space Telescope (JWST) has greatly 
expanded our ability in this regard. In particular, its NIRSpec instrument has 
been routinely discovering ${\rm H\alpha}$ emitters up to $z\approx 6$ and
[O~\Romannum{3}] emitters up to $z\approx 9$. Moreover, its unprecedented
sensitivity now allows studies in the regime more than an order of magnitude 
fainter than previously probed. Over more than three years of science operation,
JWST has accumulated a sufficiently rich spectroscopic dataset that is public to
all to explore the uncharted territory. 

   To this end, we started an investigation on ${\rm H\alpha}$ and 
[O~\Romannum{3}] emission lines using the public archival JWST NIRSpec data, 
with the simple motivation of re-examining the correlation between the two.
Past studies have explored the use of [O~\Romannum{3}] as a proxy to 
${\rm H\alpha}$ and thus a tracer for SFRs, especially at higher redshifts 
where ${\rm H\alpha}$ becomes inaccessible (e.g., the sutdy of 
\citealt{Suzuki2016} at $z=2.23$). 
\cite{Villavelez2021} found a linear relation between [O~\Romannum{3}] 
luminosity and SFR at $z\approx1.6$, although the relation shows large scatters 
due to variations in factors such as metallically. 
\cite{Wen2022} found that [O~\Romannum{3}] emission could trace star formation 
in dusty galaxies at $z=3.25$ with proper calibrations. In contrast, studies 
such as \cite{Moustakas2006} and \cite{Figueira2022} noticed that without 
accounting for different metallicities and ionization states, using the
[O~\Romannum{3}] line alone might not yield reliable SFR estimates. Despite 
these tentative results, no comprehensive studies have been done to investigate 
the correlations between  ${\rm H\alpha}$ and [O~\Romannum{3}] over a wide 
redshift range or below the limit of $\sim$$10^{-17}$~erg~s$^{-1}$~cm$^{-2}$, 
and we aimed to start filling the blank.

   Using both the low-resolution ($R\sim100$) and the mid-resolution
($R\sim1000$) spectra taken by the NIRSpec instrument, we carefully constructed 
a large sample where the ${\rm H\alpha}$ and [O~\Romannum{3}] lines could be
measured reliably. To our surprise, our initial study revealed a very tight, 
linear correlation between the ${\rm H\alpha}$ and [O~\Romannum{3}] line fluxes 
in the logarithmic space \emph{before} applying any dust reddening correction. 
While at the first glance it seems to be encouraging in that [O~\Romannum{3}] 
could be as good as ${\rm H\alpha}$ in tracing SFRs, this correlation is in fact 
problematic because dust reddening should destroy any intrinsic correlation 
between these two lines. In other words, the \emph{observed} correlation should 
not have such a linear and tight behavior with the presence of dust. Therefore, 
we went on to check the dust reddening properties of these galaxies using the
Balmer decrement method, which has long been regarded as the most reliable method
of deriving dust extinction \citep[e.g.,][]{Calzetti1994}. To our surprise 
again, we found that {$\sim$30\%} of our objects have 
${\rm H\alpha}$/${\rm H\beta}$ line ratio less
than the dust-free Case B value of 2.86. This is a more severe problem, because
the Case B recombination is the underlying assumption of the Balmer decrement
method; if the Case B assumption fails, the Balmer decrement method also fails
in measuring dust extinction. As it turns out, non-Case B Balmer decrements
have been reported sporadically in the literature; however, the universality of 
the Case B assumption was not questioned until recently
\citep[][]{Pirzkal2024, Scarlata2024, McClymont2025}. 

    While it is unclear whether they are related, these two findings are 
important enough for us to call for further investigations to understand their 
physical mechanisms. In this paper, we present them in detail.
We describe the spectroscopic data in Section~\ref{sec:data}, including the
line measurements and the sample construction. 
Section~\ref{sec:results} shows our key findings, and
Section~\ref{sec:discuss} discusses the implications. 
We summarize and conclude this work in Section~\ref{sec:conclusion}. 

\section{JWST NIRSpec Spectroscopy Data}\label{sec:data}

\subsection{Data Reduction}

We mainly used the public NIRSpec micro-shutter assembly (MSA) data taken in the 
GOODS-S, GOODS-N, and EGS 
fields for this study. These data are from the JWST Advanced Deep Extragalactic 
Survey \citep[JADES;][]{Eisenstein2023} in the two GOODS fields (PIDs 1180, 
1181, 1210, 1286, 3215; see also \citealt[][]{DEugenio24_jadesdr3}), and the 
Cosmic Evolution Early Release Science Survey \citep[CEERS;][]{Finkelstein2025} 
in the EGS field (PID 1345; see also \citealt[][]{Haro2023}). These observations 
were all done in a 3-shutter-nod configuration. The JADES data were taken in 
four disperser/filter setups: PRISM/CLEAR, G140M/F070LP, G235M/F170LP, and 
G395M/F290LP. The CEERS data were taken using similar setups, with the only 
difference that the filter used for G140M was F100LP. Hereafter, we denote the 
PRISM/CLEAR setup as ``PRISM'' and the other three grating setups as 
``Grating'', respectively. Under the PRISM setup, the nominal resolving power is 
$R\sim100$ and the spectral coverage is 0.6--5.3~$\mu{\rm m}$. The spectral 
resolution for all medium-resolution grating setups (G140M, G235M, and G395M) is 
$R\sim1000$, and the wavelength ranges are 0.70--1.27/0.97--1.84 (with 
F070LP/F100LP), 1.66--3.07, and 2.87--5.10~$\mu{\rm m}$, respectively. 

    In this work, we reduced these data on our own. We started from the Level 1b
products retrieved from the Mikulski Archive for Space Telescopes (MAST). We 
first processed them through the {\tt calwebb\_detector1} step of the standard 
JWST pipeline \citep[version 1.16.0;][]{Bushouse24_jwppl} in the context of
{\tt jwst\_1312.pmap}. We then processed the output ``rate.fits'' files through 
the {\sc msaexp} package \citep[version 0.9.2;][]{Brammar23_msaexp}, which 
provides an end-to-end reduction from the ``rate.fits'' to the final spectra
extraction. The procedure removes the ``1/f'' noise pattern and the ``snowball'' 
defects, subtracts the bias level, does the flat-fielding, applies the path-loss 
correction and flux calibration, and traces spectra on all single exposures; 
lastly, it combines all single exposures with outlier rejection. The background 
subtraction is done using the measurement in the nearest blank slit. 
{In total, we extracted 1,325 PRISM and 1,406 Grating spectra. Both the 2D 
and 1D spectra were visually inspected to make sure that the extractions were of 
good quality.
}

   For an object that has grating spectra taken in different disperser/filter 
setups, we further combined them into one single spectrum. This was done by 
first constructing a common wavelength grid that spans from the shortest to the 
longest wavelength in the individual grating spectrum. The wavelength arrays 
were then combined, and the grid adopted the sampling from the longer-wavelength 
grating in the overlapping regions. Then, each grating spectrum was resampled 
onto this common wavelength grid using linear interpolation. In regions where 
the spectra overlap, the flux values were combined using a weighted average 
based on the inverse square of their errors. 

%
%

\subsection{Sample Selection and Line Flux Measurements}

\begin{figure}
    \centering
    \includegraphics[width=0.9\linewidth]{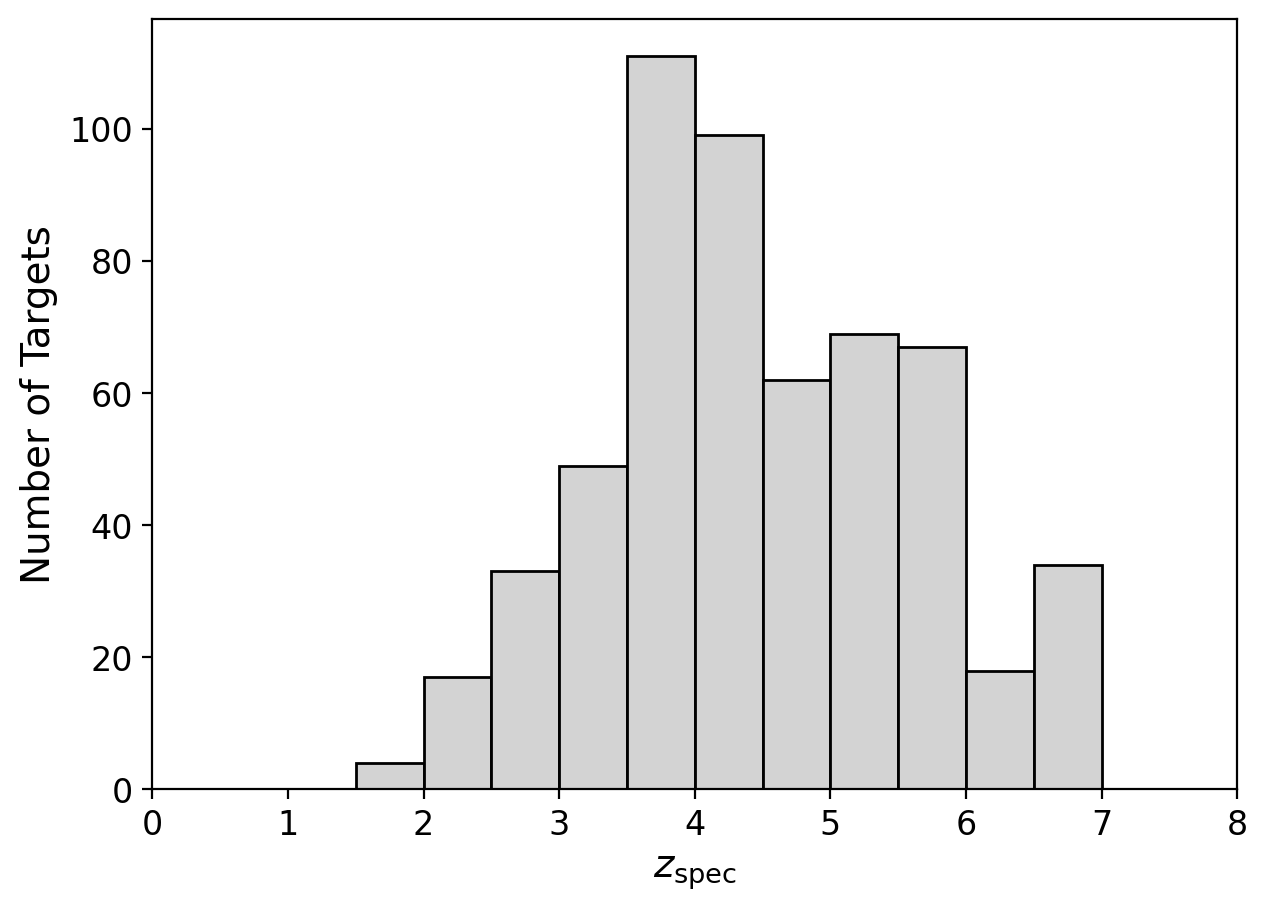}
    \caption{Redshift distribution of the 480 unique galaxies in our sample, combining both the PRISM and the Grating sets.}
    \label{fig:zspec_hist}
\end{figure}

   From the reduced data, we selected the spectra that have both ${\rm H\alpha}$ 
and [O~\Romannum{3}] detections. 
{As it turns out, 
the ${\rm H\alpha}$ emitters and the [O~\Romannum{3}] emitters are almost mutually
inclusive: both types make up $\sim$70\% of the PRISM set and $\sim$35\% of the
Grating set, respectively; only $\lesssim 2\%$ of ${\rm H\alpha}$ emitters
are not [O~\Romannum{3}] emitters, and no [O~\Romannum{3}] emitters are not
${\rm H\alpha}$ emitters.
}
We further imposed an additional selection based on the continuum fitting. 
Specifically, we performed continuum fitting on 
all available spectra using high-order polynomials (detailed below) and retained 
only those that had {good} continuum fits. This stringent selection 
minimizes the systematic uncertainties due to poor continuum subtraction 
and therefore further ensures the reliability of our line flux measurements. 
In total, 563 spectra were selected. Among these, 251 are PRISM
spectra and 312 are Grating spectra, with 83 galaxies in common. 
The total number of unique galaxies is 480. 
{The stringent selection did not introduce any preferential bias in terms of
galaxy properties: the retained galaxies have various morphologies and span a
wide range of brightnesses and line fluxes. 
}
They also span a wide redshift range of
$z\approx 1.5$--7, which is shown in Figure~\ref{fig:zspec_hist}. 
We detail our sample selection process and line flux measurements below.

\subsubsection{Low-resolution PRISM Spectra}\label{sec:prism-spec}

   We fitted the continua of the PRISM spectra with a seventh-order Chebyshev 
polynomial using the {\tt fit\_generic\_continuum} function in the 
{\sc astropy/specutils} package \citep[][]{Earl24_specutils}. The fit was
done after excluding the regions $0.1~\mu{\rm m}$ centering all strong 
emission lines. In a few cases where a seventh-order polynomial did not fit
well, we used higher-order polynomials up to the tenth order. After 
subtracting the fitted continua from the spectra, we fitted Gaussian 
profiles to the targeted lines. The line fluxes were measured within 
$2\times {\rm FWHM}$ from the central wavelengths. Most of the lines (e.g.
H$\alpha$) were fitted using a single Gaussian profile. For the
[O~\Romannum{3}]$\lambda\lambda4959,5007$ doublets, a double-Gaussian
profile was used. Some lines of different species are severely blended,
and we found that there must be at least three wavelength bins between the 
line centers in order to reliably separate them. As a result, we had to 
exclude the objects at $z<2.3$ because of the severe blending of ${\rm H\beta}$ 
and [O~\Romannum{3}]$\lambda 4959$ due to the poor spectral resolution at
$\lambda<1.65$~$\mu{\rm m}$. In the end, we selected 251 PRISM spectra that 
have both ${\rm H\alpha}$ and [O~\Romannum{3}]$\lambda\lambda4959,5007$ 
emission lines detected at SNR~$\geq 3$. We note that the line flux and
the SNR of the [O~\Romannum{3}] doublet are calculated by combining its 
two lines: {we added their fluxes as the total flux, 
and the total flux error was derived by adding the flux error of each in quadrature.}
We also note that we excluded any objects that have a broad component 
in the Balmer lines, which indicates the presence of an AGN.

\begin{figure*}
    \centering
    \includegraphics[width=0.85\linewidth]{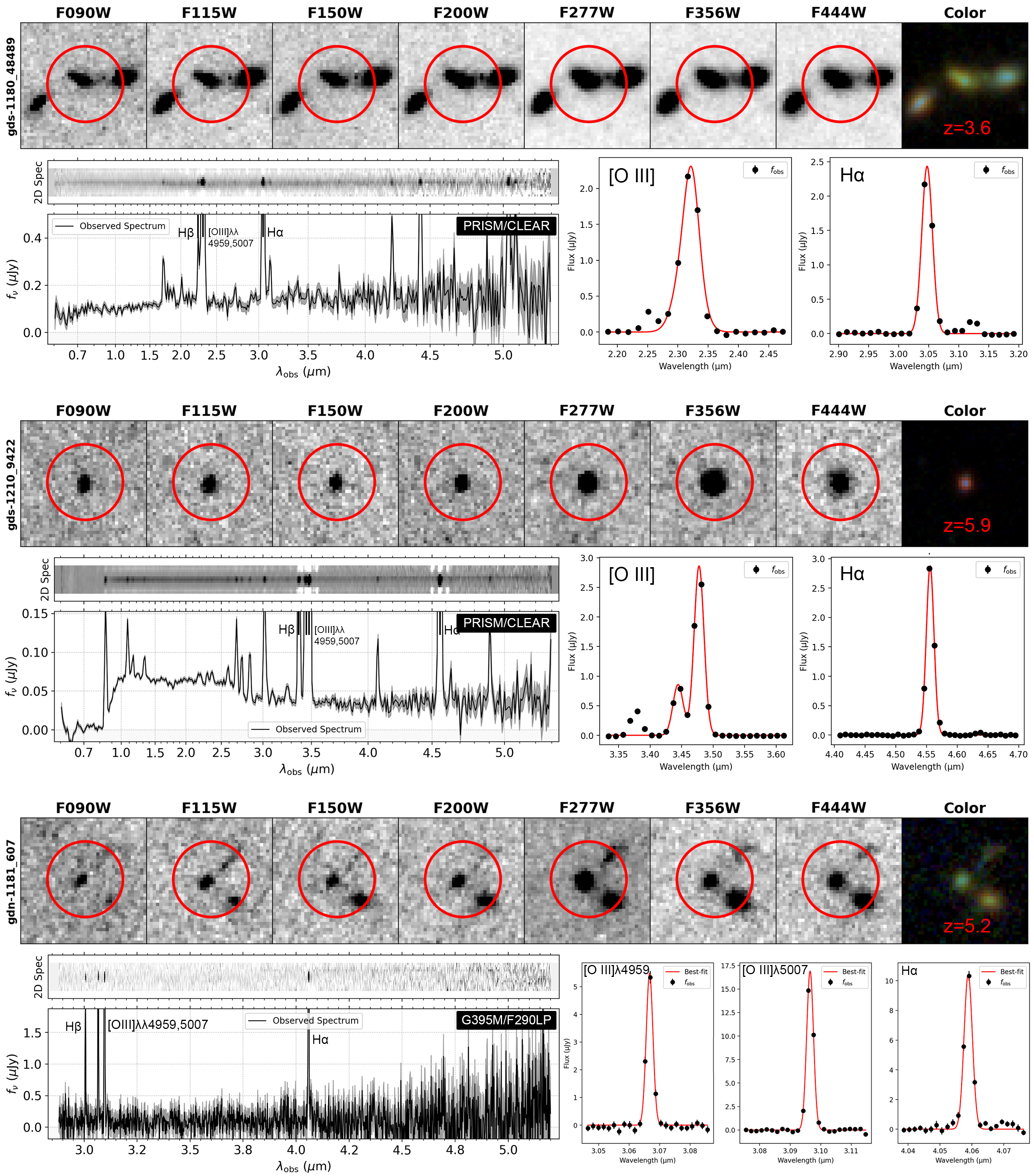}
    \caption{Example PRISM and medium-resolution grating spectra. 
    {\it Upper panel: } an object at $z=3.64$ whose [O~\Romannum{3}] doublet is unresolved in PRISM spectrum; 
    {\it Middle panel: } another PRISM spectrum whose the [O~\Romannum{3}] doublet is clearly resolved into two peaks because of its higher redshift of $z=5.94$; 
    {\it Lower panel: } an example showing individual emission line fits using the medium-resolution grating spectrum; the grating is G395M in this case. 
    }
    \label{fig:spec_example}
\end{figure*}

   Two objects from the PRISM sample are shown in
Figure~\ref{fig:spec_example} as examples. The top panel shows a galaxy
at $z=3.64$ with a blended [O~\Romannum{3}] doublet that cannot be
separated individually. As mentioned above, we fitted a double-Gaussian 
profile to the doublet to measure them as a whole.
The nearby ${\rm H\beta}$ line is shifted to
$\lambda_{\rm obs}= 2.24$~$\mu{\rm m}$, which is 
sufficiently separated from the [O~\Romannum{3}] doublet 
so that the fitting procedure could exclude ${\rm H\beta}$ 
and provide reliable measurement of the [O~\Romannum{3}] doublet.
The middle panel shows a galaxy at $z=5.94$, which has the [O~\Romannum{3}] 
doublet resolved with two clear peaks. In this case, the double-Gaussian
profile would separate the doublet.

\subsubsection{Medium-resolution Grating Spectra}\label{sec:grating-spec}

   The flux measurements and the sample selection using the grating spectra 
were done similarly to the above, with a few changes: 
(1) the continuum shapes were less complex in the grating spectra, 
and therefore we used a fifth-order polynomial instead; 
(2) all lines were fitted using a single-Gaussian profile, as the lines are
well separated;
(3) the $z\geq 2.3$ limit was lifted because the spectral resolution is 
high enough to separate ${\rm H\beta}$ from [O~\Romannum{3}] at any 
redshifts. In the end, we selected 312 grating spectra independent of the 
PRISM sample. The bottom panel of Figure~\ref{fig:spec_example} shows one 
example.

\subsection{Subsample for Balmer Decrement Measurement}\label{sec:balmerdec}

   The ``gold standard'' of deriving dust reddening of galaxies is to use 
the Balmer decrement, most commonly the ratio between ${\rm H\alpha}$ and
${\rm H\beta}$, under the Case B assumption of hydrogen recombination.
A fraction of our objects have strong enough H$\beta$ line, allowing us to 
carry out this practice. To this end, we selected a subsample where the
${\rm H\beta}$ line detections have SNR~$\geq 3$. This subsample consists 
of 179 PRISM and 265 grating spectra, with 66 in common. While selecting 
this subsample, we also searched for cases where both Balmer lines are 
detected but [O~\Romannum{3}] is not; we did not find such a case among 
all the spectra that we reduced. 
This means that our subsample for the Balmer 
decrement measurement is not biased in terms of the [O~\Romannum{3}] line
strength.

    In what follows, we will derive the gas-phase dust reddening by following 
the extinction law of \citet[][]{Calzetti2000}:
\begin{equation}\label{eq:ebv}
    E(B-V)_{\rm gas}=\frac{\log_{10}(R_{\rm obs}/R_{\rm int})}{0.4(k_{\rm H\beta}-k_{\rm H\alpha})},
\end{equation}
where $k_{\rm H\alpha}=3.327$ and $k_{\rm H\beta}=4.598$. $R_{\rm obs}$ is the 
observed line flux ratio between ${\rm H\alpha}$ and ${\rm H\beta}$, and 
$R_{\rm int}=2.86$ is the widely adopted intrinsic ratio in the Case B 
recombination at an electron temperature of $T_e=10^4$~K and an electron density
of $n_e=10^2$~cm$^{-3}$ \citep[][]{Osterbrock1989}.

\subsection{Consistency Check of Line Measurements in PRISM and Grating Sets}

\begin{figure*}[hbt!]
    \centering
    \includegraphics[width=0.85\linewidth]{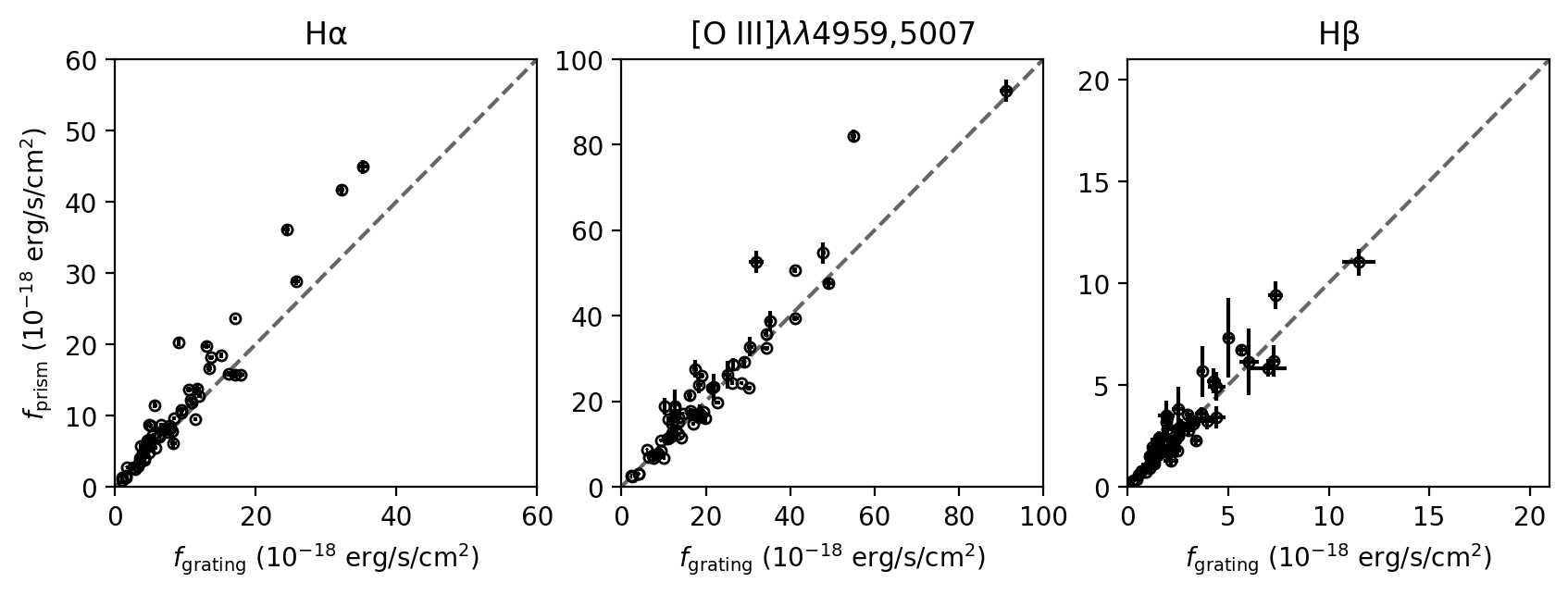}
    \caption{
    Comparison of line fluxes for ${\rm H\alpha}$ (left), [O~\Romannum{3}] 
    (center), and ${\rm H\beta}$ (right) measured from the PRISM and the Grating 
    spectra for the common objects. The X-axis represents the flux measurements 
    from the Grating set ($f_{\rm grating}$) and the Y-axis represents those from 
    PRISM set ($f_{\rm prism}$), both in the unit of $10^{-18}~{\rm erg/s/cm^2}$. 
    A straight line is plotted to indicate 1:1 agreement between the two 
    measurements. 
    }
    \label{fig:prism-grating-consistency}
\end{figure*}

   To ensure that we can use a combined sample of the PRISM and the Grating sets,
we conducted a consistency check of the line flux measurements on
${\rm H\alpha}$, [O~\Romannum{3}], and ${\rm H\beta}$ emission lines uisng the 
common objects in these two sets. This check is presented in 
Figure~\ref{fig:prism-grating-consistency}. 

   {The fluxes generally agree well, although there are small systematics 
present. The most significant (albeit still small) offset is seen in 
${\rm H\alpha}$, and there are cases where the flux measurements from the PRISM 
spectra are notably stronger. A simple linear fit to the left panel in 
Figure~\ref{fig:prism-grating-consistency}
gave the best-fit slope of $\sim$1.05, i.e., the $\rm H\alpha$ fluxes measured 
from the PRISM set are approximately 5\% stronger than those measured from the 
grating spectra. The simplest explanation would be that the low resolution 
of the PRISM spectra causes the blending of the nearby [N~\Romannum{2}] 
lines and therefore results in an overestimate of the ${\rm H\alpha}$ flux.} 
{To verify this, we further investigated the [N~\Romannum{2}] line. Among the 
83 common galaxies between the two sets, 54 have their $\rm H\alpha$ fluxes 
measured from the PRISM spectra being $\geq 1\sigma$ higher than those measured 
from the grating spectra, and 91\% of them show [N~\Romannum{2}]$\lambda$6583 
detection in the latter. 
We measured the [N~\Romannum{2}]$\lambda$6583 line fluxes and calculated the 
[N~\Romannum{2}]/$\rm H\alpha$ line ratios. The median value of the ratio is
6.9\%, which is in broad agreement with the aforementioned $\sim$5\% systematic 
offset. 
}
{In short, we conclude that the small differences between the PRISM and
the Grating spectra do not significantly affect our analysis on the combined 
sample.}

\section{Results}\label{sec:results}

\subsection{Tight Correlation between ${\rm H\alpha}$ and [O~\Romannum{3}]} \label{sec:tight_relation}

\begin{figure*}
    \centering
    \includegraphics[width=0.49\linewidth]{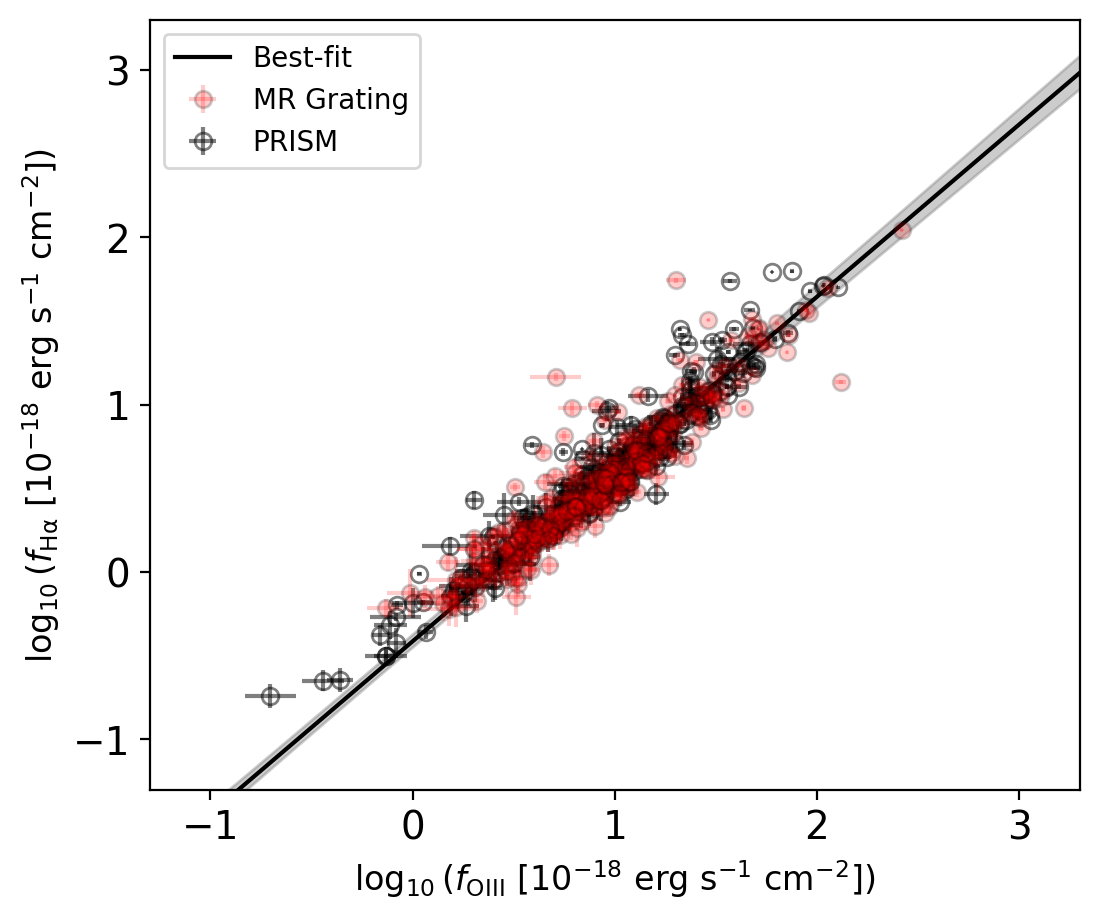}
    \includegraphics[width=0.49\linewidth]{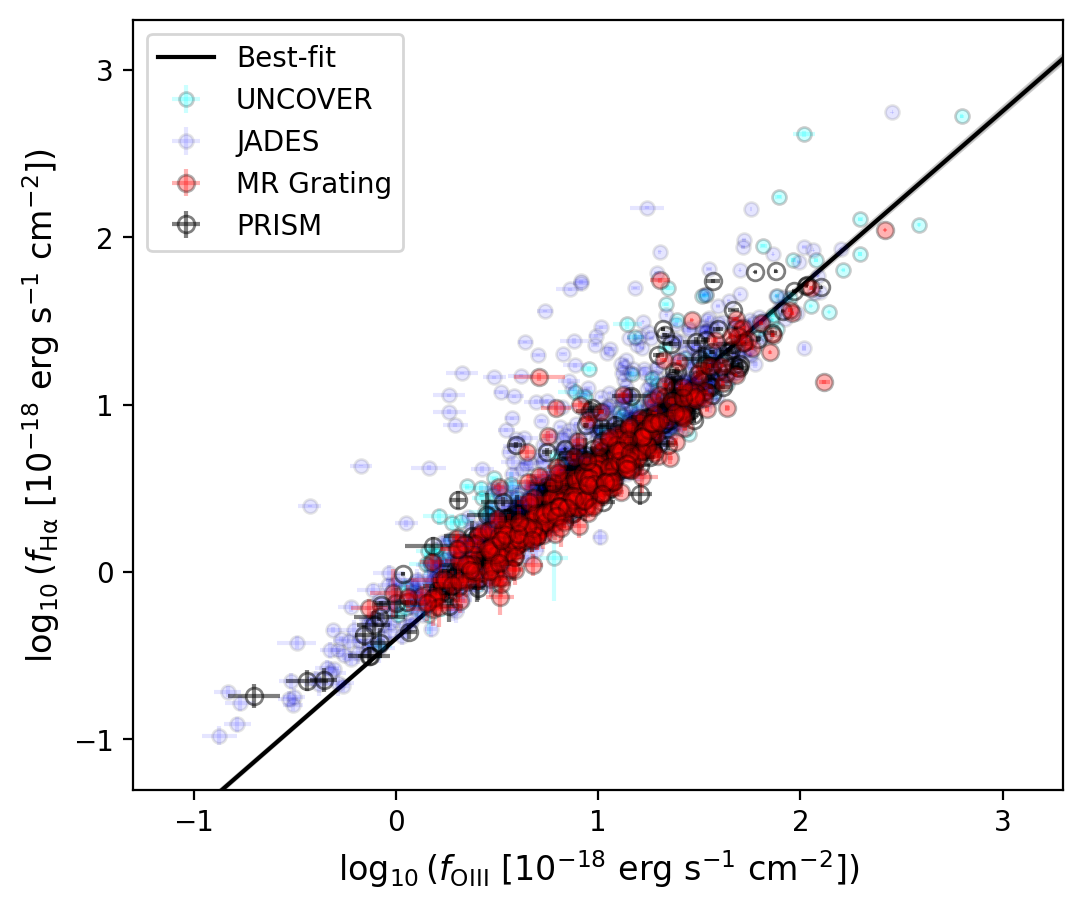}\\
    \includegraphics[width=0.95\linewidth]{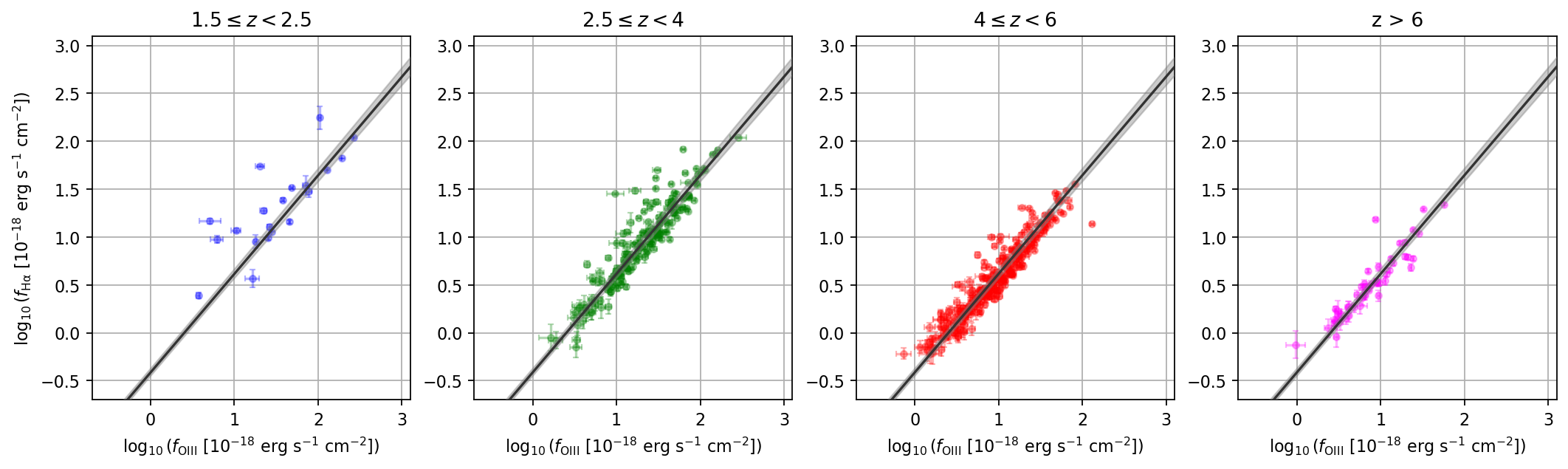}\\
    \includegraphics[width=0.95\linewidth]{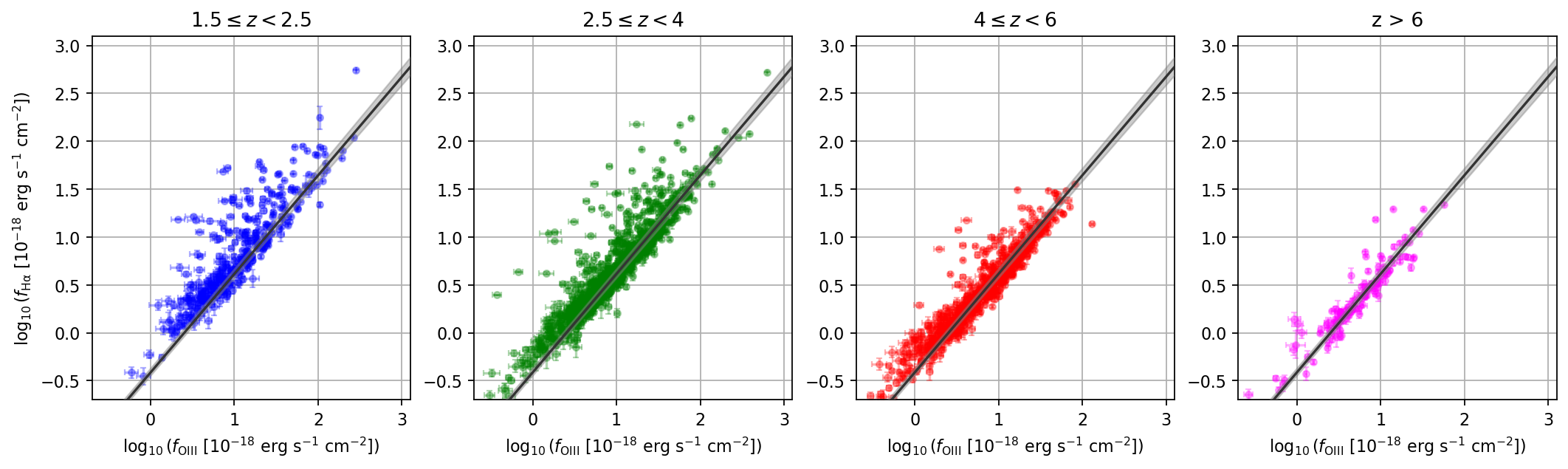}
    \caption{{\it Top panels: } Tight linear relation between ${\rm H\alpha}$ and
     [O~\Romannum{3}] line fluxes in logarithm as revealed by our sample (left)
     and also seen in the NIRSpec catalogs released by the JADES and the UNCOVER
     teams (right). The straight line
     in the left panel is the best-fit to the combined PRISM and Grating sets,
     while the one in the right panel is the best-fit to our sample plus the
     ones from the JADES and the UNCOVER catalogs. The 2~$\sigma$ uncertainty is 
     indicated by the gray areas. 
    {\it Middle panels: } Similar to the top-left panel (i.e., using only the
    data from our sample) but to show the ${\rm H\alpha}$--[O~\Romannum{3}]
    correlation in four successive redshift bins. 
    {\it Bottom panels: } Similar to the middle panels but using the data combining 
    our sample and those from JADES and UNCOVER programs.
    }

    \label{fig:line_relations}
\end{figure*}

   The left panel of Figure~\ref{fig:line_relations} shows the relation 
between the ${\rm H\alpha}$ and the [O~\Romannum{3}] doublet line fluxes in 
logarithm, using both the PRISM sample (black symbols) and the Grating
sample (red symbols). Surprisingly, the relation is an extremely tight
linear correlation. We fitted a simple linear function in the form of 
\begin{equation}\label{eq:line_relations}
    \log_{10}(f_{\rm H\alpha})=k\times\log_{10}(f_{\rm [O\Romannum{3}]})-b
\end{equation}
and obtained the best-fit coefficients $(k,b)=(1.03\pm0.03,-0.41\pm0.04)$, 
with the root mean square (rms) value of only 0.104. 
This linear form is shown as the solid black line in the figure, with 
the gray-shaded regions indicating the 2~$\sigma$ error range. Separating
the PRISM and the Grating samples does not make any notable difference in
the fitted relation.

   Such a tight linear correlation is unexpected, as not all the galaxies can 
be dust-free and therefore statistically the [O~\Romannum{3}] line must 
suffer more severe extinction than the ${\rm H\alpha}$ line does. To verify
whether this could be due to our sample being somewhat biased {for unknown
reasons}, we checked for the correlation
using the results from two other teams, namely, the spectroscopic catalogs from 
the public releases of the JADES DR3 \citep[][]{DEugenio24_jadesdr3} and the DR4 
of the Ultradeep NIRSpec and NIRCam Observations before the Epoch of 
Reionization program 
\citep[UNCOVER;][]{Bezanson2024_uncover,Price2024_uncover-spec}. Both catalogs
are based on the PRISM spectra. 
We only used the objects with reliable spectroscopic redshifts 
$z_{\rm spec}$ from these catalogs (ranks A, B, C in JADES DR3; ranks 2, 3 
in UNCOVER) and only retained those that have reported ${\rm H\alpha}$ and 
[O~\Romannum{3}] detections at SNR~$\geq 3$ to be consistent with our 
sample selection (see Section 2.2.1). In total, there were 1128 and 164 spectra
from JADES and UNCOVER used in the comparison, respectively. The correlation
between ${\rm H\alpha}$ and [O~\Romannum{3}] from these independent measurements 
of two different teams is the same as ours, and the comparison is shown in the 
right panel of Figure~\ref{fig:line_relations}. 
There are 428 objects in common between our sample and the JADES sample. Had
there been any systematic differences between these two sets, one
would observe some offsets for these common objects in this comparison.
However, no such offsets are seen. Therefore, we conclude that our sample is not
biased, nor are our line measurements erroneous. 
Fitting the same linear relation with all three datasets combined
(ours + JADES + UNCOVER), 
we got $(k,b)=(1.05\pm0.02,-0.37\pm0.01)$ and RMS value of 0.160; in other words,
both the slope and intercept are only marginally different from what we
obtained using only our sample. 
When fitting the same relation with JADES and UNCOVER alone, 
we got $(k,b)=(1.07\pm0.02,-0.43\pm0.02)$ and $(1.10\pm0.04,-0.41\pm0.06)$, 
and RMS values of 0.150 and 0.143,
respectively. 


    To investigate the possible redshift dependency of the correlation,
we divided the spectra into four redshift bins: $1.5\leq z<2.5$, $2.5\leq z<4$, 
$4\leq z<6$, and $z\geq6$. Figure~\ref{fig:line_relations} show the 
${\rm H\alpha}$--[O~\Romannum{3}] relation in these four bins using our sample
alone (middle row) and when combining all the three samples (bottom row).
In all four redshift bins, the Pearson correlation coefficient consistently 
remains $R>0.8$, and notably, the correlation appears to be even tighter at 
higher redshifts. We fitted the same linear relation in these four ranges, 
and the best-fits to our sample have $(k,b)=(0.90\pm0.09,-0.21\pm0.16)$, 
$(1.06\pm0.02,-0.49\pm0.04)$, $(0.97\pm0.02,-0.39\pm0.02)$, and
$(0.96\pm0.04,-0.36\pm0.05)$, respectively; 
and those to three datasets combined are
$(k,b)=(1.11\pm0.02,-0.36\pm0.02)$, $(1.02\pm0.02,-0.37\pm0.02)$, 
$(0.95\pm0.02,-0.34\pm0.03)$, $(0.97\pm0.03,-0.40\pm0.03)$. 
While these values seem to suggest a slight decrease in the slope at 
higher redshifts, the differences are not statistically significant. If there
is any redshift-dependent trend, it is that the relation seems to be tighter at
higher redshifts.



\subsection{Comparing to the pre-JWST results}\label{sec:prejwst}

\begin{figure*}
    \centering
    \includegraphics[width=0.95\linewidth]{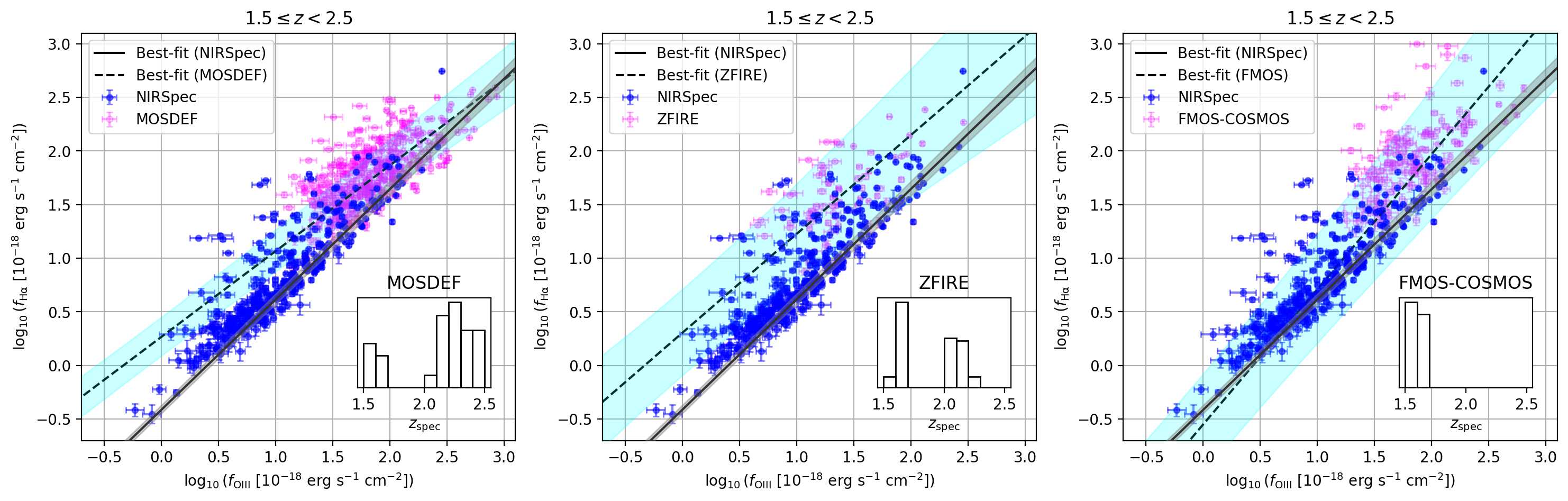}
    \caption{Comparison of the ${\rm H\alpha}$--[O~\Romannum{3}] relation 
    obtained with NIRSpec (combining our sample, JADES and UNCOVER; blue symbols) and that from the three ground-based surveys
    (magenta symbols) using Keck/MOSFIRE (the MOSDEF and zFIRE programs) and 
    Subaru/FMOS (the FMOS-COSMOS program) in our lowest redshift bin of 
    $1.5\leq z< 2.5$. The solid black straight line and the gray areas indicate
    the same linear correlation and the dispersion as in the lower-left panel of 
    Figure~\ref{fig:line_relations}, while the dashed straight line and the cyan
    areas indicate the best linear fit to the three ground-based survey data and 
    the dispersion. The insets show the redshift distributions of the three
    ground-based samples.
    }
    \label{fig:nrs_mosdef_compare}
\end{figure*}

    Surprised by the tight correlation discussed above, we checked the published 
results from a few large spectroscopic surveys in the pre-JWST era. As it turns 
out, such a correlation is also present in these data, albeit with much larger 
dispersions. Here we highlight the comparison with the measurements from the 
MOSFIRE Deep Evolution Field program \citep[MOSDEF;][]{Kriek15_mosdef} and the 
ZFIRE program \citep[][]{Nanayakkara2016} carried out by the MOSFIRE instrument 
at the Keck-1 telescope and the FMOS-COSMOS program \citep[][]{Kashino2019} done 
by the FMOS spectrograph at the Subaru telescope. These three programs spanned 
the redshift range of $z\sim1$--2.5, for which we can make a direct comparison. 
This is demonstrated in Figure~\ref{fig:nrs_mosdef_compare}, where the NIRSpec
data points in this range are shown in blue and those from the three 
ground-based programs are shown in magenta. The latter also follows a linear trend
with $(k,b)=(0.80\pm0.06, 0.32\pm0.15)$, 
but the distribution lies slightly above our relation
and has a much larger dispersion (RMS of 0.403). 
    We also checked the line flux measurements from the Sloan Digital Sky Survey 
(SDSS) value-added MPA-JHU catalog
\footnote{\href{https://wwwmpa.mpa-garching.mpg.de/SDSS/DR7/}{https://wwwmpa.mpa-garching.mpg.de/SDSS/DR7/}}
for the situation at $z<1$. The correlation is no longer obvious in these data. 
As currently there are insufficient number of NIRSpec measurements at $z<1$, we 
do not make a direct comparison here.

\subsection{Conflict with Dust Reddening}

    The very tight, linear correlation between $\log_{10}({\rm H\alpha)}$ and
$\log_{10}({\rm [O~\Romannum{3}]})$ seen in the NIRSpec spectra suggests that 
there is an intrinsic, fixed ratio between the two lines. However, even if such 
a fixed ratio exists, the observed ratios would deviate from this value because 
of the dust reddening. In other words, such a relation would only make sense if, 
for some reason, the target selections for NIRSpec spectroscopy by these 
programs were all biased in favor of dust-free galaxies. However, this does not 
seem to be the case.

    Our subsample for the Balmer decrement measurement allows us to derive the 
gas-phase dust reddening ${\rm E(B-V)_{gas}}$ for these objects (see 
Section~\ref{sec:balmerdec}). Under the standard Case B assumption, most of them 
have significant reddening, which is inconsistent with the linear relation 
between $\log_{10}({\rm H\alpha)}$ and $\log_{10}({\rm [O~\Romannum{3}]})$: if the
intrinsic relation between the two is indeed linear, dust reddening would
suppress {\rm [O~\Romannum{3}]} more than ${\rm H\alpha}$, which would then
skew the linear relation. In other words, the dust reddening and the 
intrinsic line ratio would have to conspire in a delicate way such that the 
observed relation is linear. This point is demonstrated in the top-left panel of 
Figure~\ref{fig:ratio_ebv} in the form of flux ratio of ${\rm H\alpha}$ and 
${\rm [O~\Romannum{3}]}$ versus ${\rm E(B-V)_{gas}}$. As $k$ in 
Equation~\ref{eq:line_relations} is very close to unity, the line ratio
represents well the observed log-linear relation. The data points concentrate 
at the value of $f_{\rm H\alpha}/f_{\rm [OIII]}\approx0.4$, which is the 
manifest of the best-fit intercept of $-0.41$ that we obtained 
($10^{-0.41}\approx 0.39$) in Figure~\ref{fig:line_relations}.
If there is an intrinsic, fixed ratio between the two lines, the observed line
ratio should be changed by the dust reddening. This is indicated by the yellow 
dashed curve, which is the predicted behavior of the observed line ratio due to
reddening when assuming an intrinsic ratio of 0.4. The top-right panel shows the 
multiplicative factor that needs to be applied to move the data points to the 
yellow dashed curve, which are effectively the ratios of the observed and the
predicted $f{\rm (H\alpha)}/f([{\rm O~\Romannum{3}}])$. 

\subsection{Non-Case B ${\rm H\alpha}$/${\rm H\beta}$ Ratios}

    A more severe problem is revealed while investigating the dust reddening. As
shown in these two panels, there are a large number of objects that have
negative ${\rm E(B-V)}$ values (see the grey area), which is unphysical. We find
that this is due to a large number of objects having
$f_{\rm H\alpha}/f_{\rm H\beta}$ values less than the canonical Case B value 
of 2.86. 
{The fraction of such cases is $\sim$30\% in both the PRISM and 
the Grating sets. To investigate whether the weaker $\rm H\beta$ line would bias this 
ratio low, we carried out a Monte-Carlo simulation, which is detailed in
Appendix~\ref{appendix:monte-carlo}. Basically, there is no bias.}

\begin{figure*}
    \centering
    \includegraphics[width=0.48\linewidth]{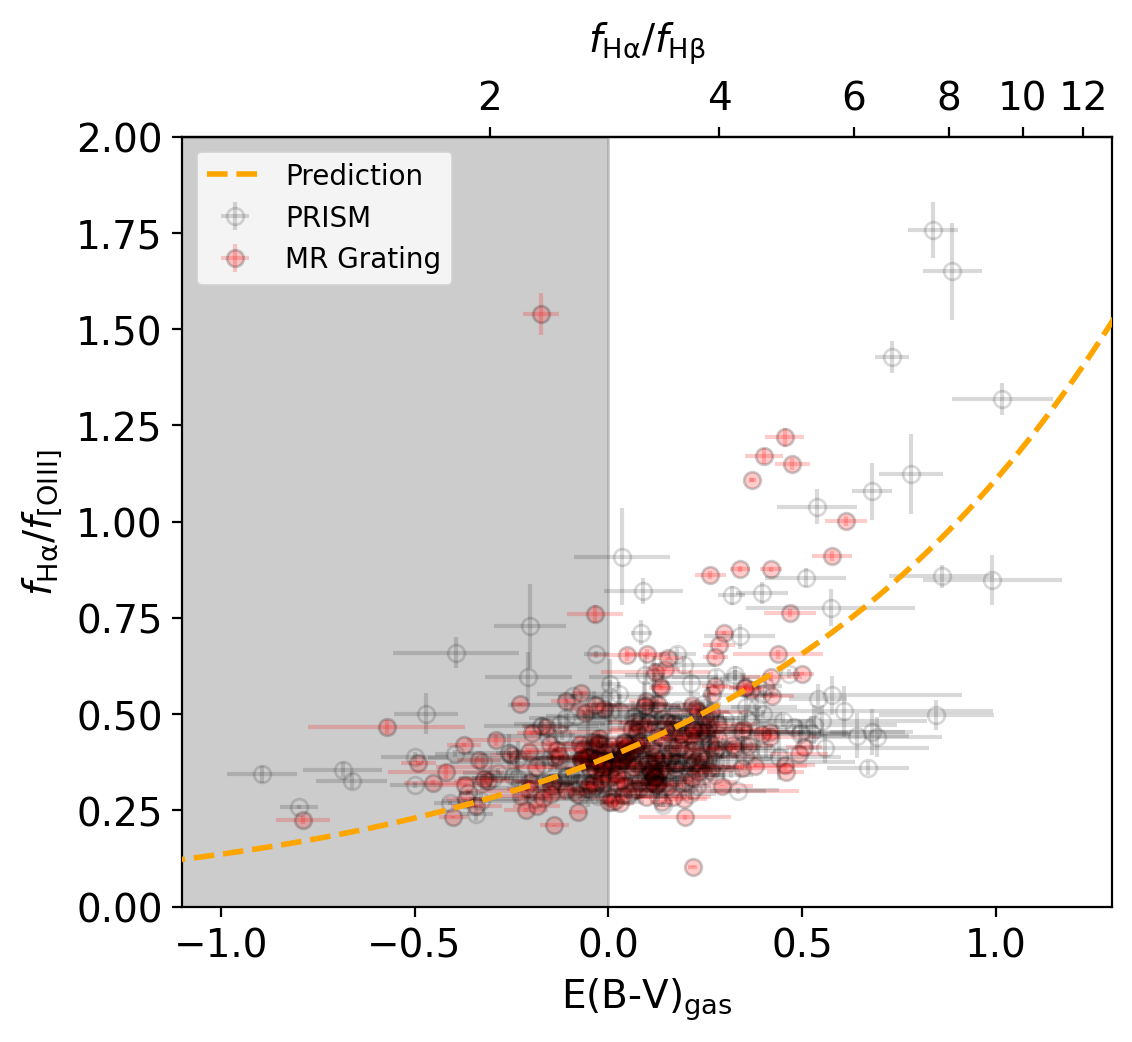}
    \includegraphics[width=0.47\linewidth]{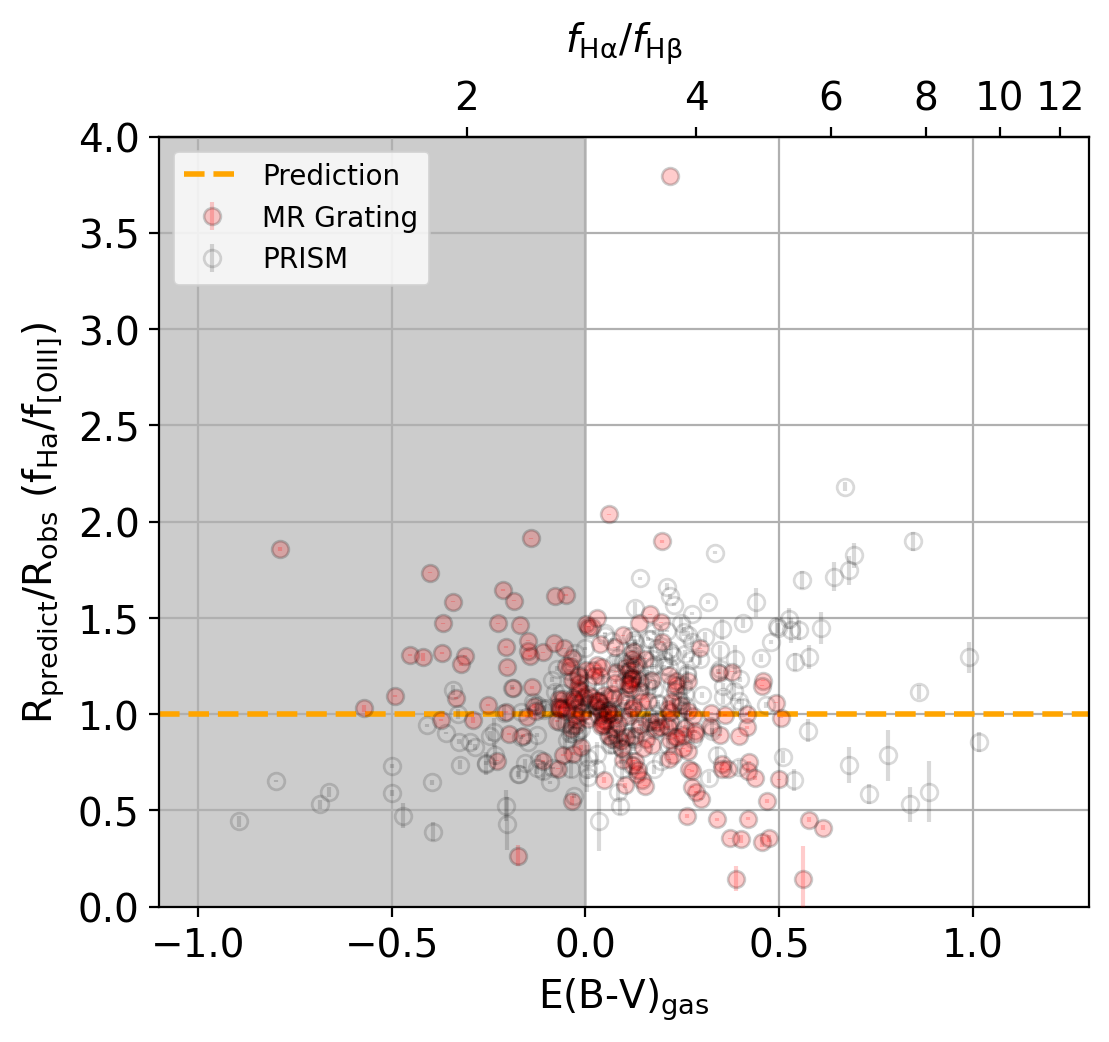} 
    \includegraphics[width=0.9\linewidth]{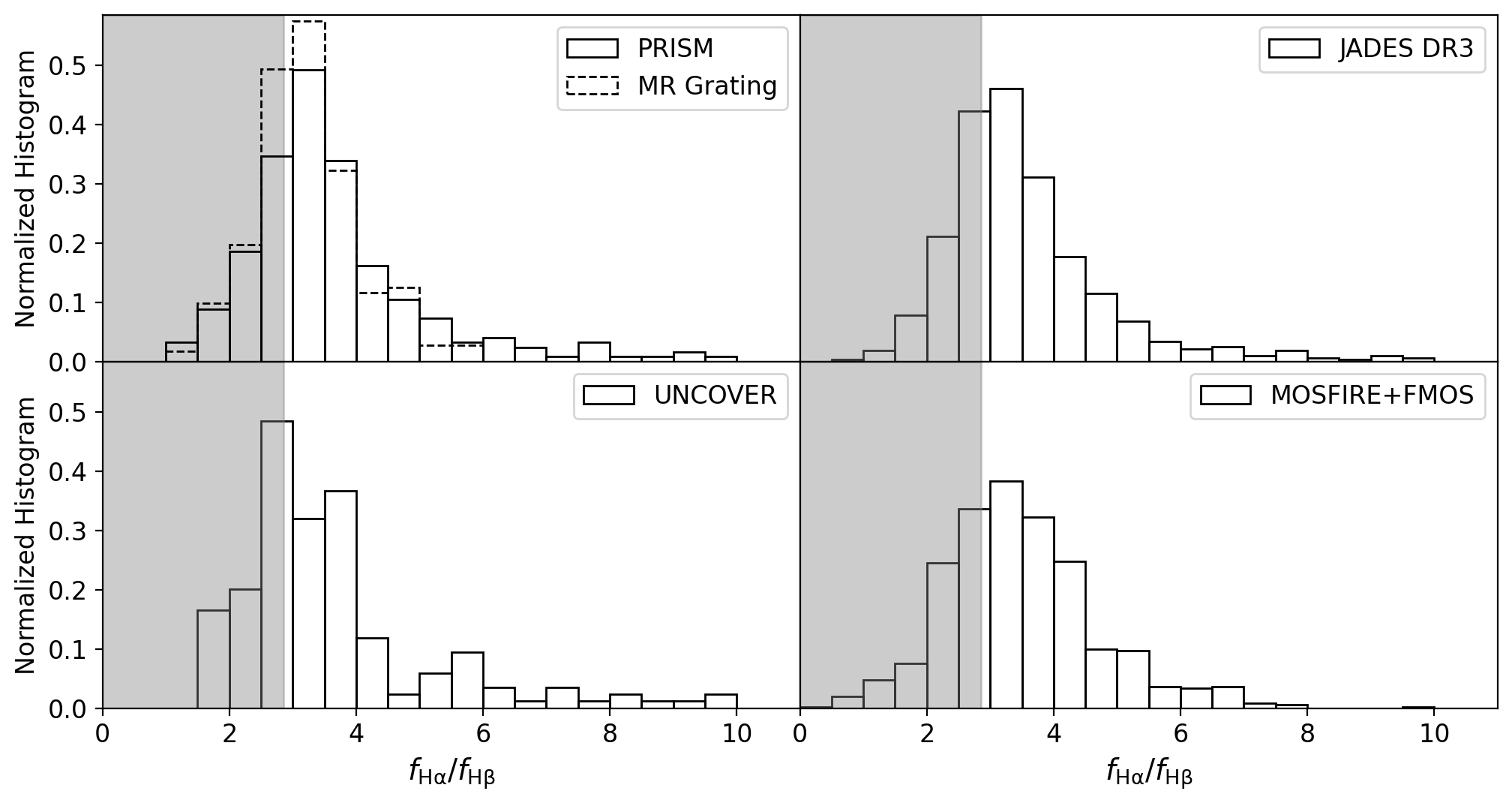}
\\ 
    \caption{
    {\it Top-left: } Relations between the ${\rm H\alpha/}$[O~\Romannum{3}] flux 
    ratios and the estimated gas-phase dust reddening ${\rm E(B-V)_{gas}}$ under 
    the Case B assumption and the Calzetti's extinction law. The PRISM and the 
    Grating data from our sample are shown as the gray and the red symbols, 
    respectively. The dashed orange curve is the predicted behavior of 
    ${\rm H\alpha/}$[O~\Romannum{3}] as a function of ${\rm E(B-V)_{gas}}$. The
    gray-out area is where ${\rm E(B-V)_{gas}<0}$ due to the non-Case B ratios
    (see the bottom panels).
    {\it Top-right: } Multiplicative factors that need to be applied to the 
    observed ${\rm H\alpha/}$[O~\Romannum{3}] ratios to move them to the 
    predicted behavior as shown in the left panel.
    {\it Bottom: } Histogram of ${\rm H\alpha/H\beta}$ line ratios from our 
    sample, from two other public catalogs using JWST NIRSpec (JADES and 
    UNCOVER) and from three ground-based programs (MOSDEF, zFIRE, and 
    FMOS-COSMOS), respectively. The line ratios smaller than the canonical 
    Case B values of 2.86 (indicated by the gray-out area) are present in high 
    fraction ($\sim$30\% in our sample and  $\sim$30\% in others) in all these 
    samples, suggesting the Case B assumption is not universally applicable.
    }
    \label{fig:ratio_ebv}
\end{figure*}

     Being surprised, we check this line ratio in the JADES and the 
UNCOVER catalogs as well as in the three ground-based redshift survey catalogs 
discussed in Section~\ref{sec:prejwst}. As it turns out, all
these catalogs have a large fraction of objects ($\sim$30\% in JADES and UNCOVER 
and also $\sim$30\% in MOSFIRE+FMOS) that have 
$f_{\rm H\alpha}/f_{\rm H\beta}<2.86$, which is shown in the histograms in 
Figure~\ref{fig:ratio_ebv}. 
{The non-Case B ratios in the JADES catalog were also recently
investigated by \citet[][]{McClymont2025}, who found a similar result.}

    {To assess how robust the non-Case B samples remain when the 
measurement errors are considered, we also calculated the percentage of galaxies 
with their Balmer decrement upper limit $R_{ul}$ still falling below the 
canonical dust-free value of 2.86. Specifically, we required 
$R_{ul}=R+n\times {\rm Err_R}<2.86$ for $n=1,2,3$, where $n\times {\rm Err_R}$ is the 
$n\sigma$ error in $R$. The results are summarized in 
Table~\ref{tab:non-caseb-percentage}. In most samples, the non-Case B percentage
still maintains at $\sim$15-20\% after allowing the $1\sigma$ uncertainty; at 
even more conservative $2\sigma$ and $3\sigma$ levels the fractions are still at 
least $\sim$3-5\%. In other words, at least about a dozen galaxies in our sample 
are still inconsistent with the Case B recombination under the generous allowance
of 3~$\sigma$ error, and for other samples of larger sizes (e.g., JADES and
UNCOVER) there are even more such galaxies. This shows that the non-Case B Balmer 
decrements cannot be explained simply by measurement errors; instead, 
}
this is a strong indication that the Case B assumption is not universally 
applicable, which we will discuss in the next section. 

\begin{table}[hbt!]
    \raggedright
    \scriptsize
    \caption{{Percentage of galaxies whose Balmer decrement 
    upper limit $R_{\rm ul}$ falls below the canonical 
    Case B value of 2.86 at different $\sigma$ levels in $R$. }
    }
    \begin{tabular}{cc}
        Sample & \%$R_{\rm ul}\leq 2.86$ at $n=0,1,2,3$ \\ \hline 
        {\bf Ours} & 31,13,6,5 \\ 
        JADES & 29,15,7,3 \\ 
        UNCOVER & 31,20,9,4 \\ \hline 
        MOSDEF & 26,14,6,4 \\ 
        ZFIRE & 32,19,15,15 \\ 
        FMOS-COSMOS & 44,32,20,14 \\ \hline 
    \end{tabular}
    \tablecomments{The upper limit is defined as $R_{\rm ul}=R+n\times {\rm Err_R}$ 
    where $R$ is the line flux ratio between ${\rm H\alpha}$ and ${\rm H\beta}$. 
    }
    \label{tab:non-caseb-percentage}
\end{table}

\section{Discussion}\label{sec:discuss}

\subsection{Path-loss Correction in NIRSpec MSA Spectra} 

Path loss, or slit loss, is a universal problem in slit spectroscopy when 
the slit is not wide enough to cover the target entirely. The micro-shutter 
apertures of the NIRSpec MSA
are small at a width of only 0\farcs20. Therefore, a fraction of a target's light 
may fall outside of its shutter. For an extended source that is only partially 
covered by the shutter, this effect can be largely corrected by scaling the 
spectrum to the proper photometry of the entire source. As our study relies 
solely on line ratios, this effect is not a concern. However, there is a more 
subtle secondary effect, which is wavelength-dependent and should be discussed 
for our study. As the NIRSpec PSF size varies with wavelength (optimized at
$\sim$1.5~$\mu$m), the fraction of the lost light also varies with wavelength.
The {\sc msaexp} algorithm remedies this problem empirically. For every 
extracted spectrum, {\sc msaexp} traces it and calculates the path-loss at each 
wavelength bin based on the width of the fitted profile and the expected 
intra-shutter position. This process uses the empirical data taken within the 
slit and therefore has already considered the PSF variation with wavelength
as well as the source position in shutter. Therefore, our analysis is not likely
to be affected by this secondary slit-loss effect. 
{The path loss correction factor is applied to both the science and error 
extensions; the uncertainties due to slit-loss correction are already 
propagated into the final, extracted 1D spectrum. 
}


\subsection{Comparison of Dust Extinction Derived from Two Methods}

    To further investigate the dust reddening problem, we used a different
approach. For all the galaxies in our subsample for measuring the Balmer 
decrements, we derived their dust extinction
$A_{\rm V}^s$ by fitting their spectral energy distributions (SEDs), which can
then be compared to the gas-phase dust extinction $A_{\rm V}^g$ based on the 
Balmer decrements discussed above. Note that $A_{\rm V}^s$ is the extinction
of stellar light and is different from the gas-phase extinction; adopting the 
Calzetti's extinction law implies that $A_{\rm V}\approx 4.04\times E{\rm (B-V)}$
and $E{\rm (B-V)_{star}}=0.44\times E{\rm (B-V)_{gas}}$.

    The SEDs of all our objects were constructed using the PSF-matched photometry
of \citet[][]{SY2025a}. To fit the SEDs, we used {\sc Bagpipes} 
\citep[][version 1.2.0]{Carnall18_bagpipes}, which utilizes the stellar 
population synthesis models of \citet[][]{Bruzual2003} with the initial mass
function of \citet[][]{Kroupa2001}. We ran the program using the 
delayed-$\tau$ star formation history, which is in the form of 
${\rm SFR}\propto te^{(-t/\tau)}$. 
We set $\tau$ as a free parameter that could vary between $0.01$--15~Gyr. 
The metallicity was also set free in the range of $0\leq Z/Z_\odot \leq2.5$, 
and the dust extinction $A_V$ could vary from 0 to 8 mag. 
We also enabled the option to include nebular emission lines, 
and the ionization parameter could vary between $-4.0\leq \log(U)\leq -2.0$. 
The redshift of each target is fixed at its $z_{\rm spec}$. 
   
\begin{figure}
    \centering
    \includegraphics[width=0.9\linewidth]{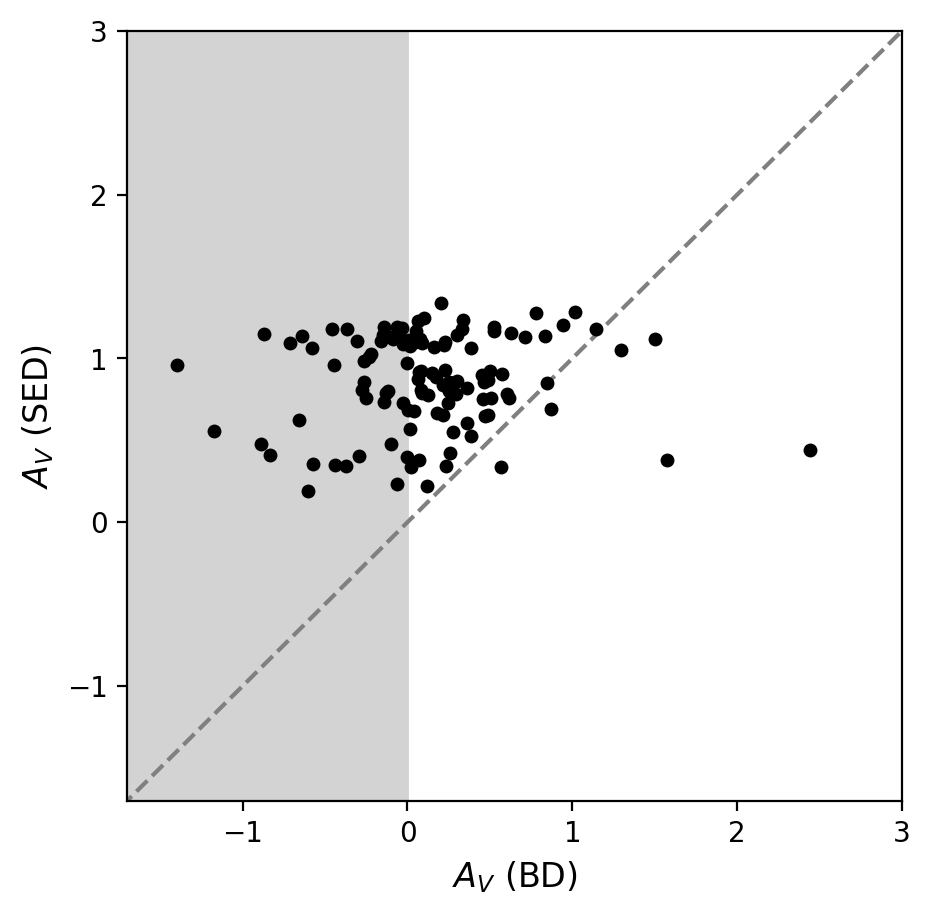}
    \caption{Comparison between dust extinction $A_V$ derived from the SED fitting 
    (denoted as $A{\rm _V(SED)}$) and converted from the gas-phase extinction 
    based on the Balmer decrement (denoted as $A{\rm _V(BD)}$).
    The dashed black line shows the 1:1 correspondence. The grey-out area is 
    where the objects have $A{\rm _V(BD)}<0$ due to their ${\rm H\alpha/H\beta}$
    line ratio being lower than the canonical Case B value of 2.86. 
    There is hardly any trend between the two, which further enhances the
    argument that the Case B assumption is not universally applicable.
    }
    \label{fig:av_balmer_sed}
\end{figure}

   For clarity, we denote the $A_{\rm V}^s$ values derived from the SED fitting 
above as $A{\rm _V(SED)}$ and the $A_{\rm V}^s$ values converted from 
$E{\rm (B-V)_{gas}}$ based on the Balmer decrement measurements as 
$A{\rm _V(BD)}$. The comparison of the two is presented in    
Figure~\ref{fig:av_balmer_sed}. Simply put, the two do not agree, and
$A{\rm _V(SED)}$ is larger than $A{\rm _V(BD)}$ in most cases. To some extent,
the disagreement could still be explained if the star-forming regions that give
rise to the nebular emission lines are not well mixed with the rest of the
stellar populations. However, it is more likely that this is another indication
of the failure of measuring dust extinction using the Balmer decrement. In
particular, those with unphysical $A{\rm _V(BD)}<0$ still have 
significant $A{\rm _V(SED)}$.


\subsection{Non-Case B Recombination}

    All our analysis above suggests that the usual practice of deriving dust
extinction through the Balmer decrement can be problematic. The strongest
indication comes from the high fraction of galaxies that have ${\rm H\alpha}$
and ${\rm H\beta}$ line ratios smaller than the dust-free Case B value.
Such objects are present in both
our PRISM and Grating sets and therefore cannot be attributed to different 
spectral resolutions of the data. They are also found in high fractions in at 
least two public NIRSpec catalogs from two independent teams and therefore are 
not likely due to measurement flaws. Moreover, they also exist in similarly
high fractions in three ground-based redshift survey programs in the pre-JWST
era, which means that it is a universal problem already observed but
largely unexplored. In fact, the SDSS value-added MPA-JHU catalog also contains 
such objects at a low but non-negligible fraction ($\sim$6\%), however this was
not discussed in the literature until recently \citep[see][]{Scarlata2024}. 
There have been a few studies that noticed such anomalies but did not discuss
them in the context of non-Case B possibility, which we summarize in 
Appendix~\ref{appendix:literature}.

     The universality of the Case B assumption was recently questioned by
\citet[][]{Pirzkal2024}, who analyzed the grism slitless spectroscopic data
($R\sim 150$) taken by the JWST NIRISS instrument in the historic Hubble Ultra 
Deep Field. Among their 91 galaxies at $1<z<3.5$ whose ${\rm H\alpha}$ and 
${\rm H\beta}$ lines are detected at ${\rm SNR>2}$, 30\% have the 
${\rm H\alpha}$/${\rm H\beta}$ line ratios smaller than 2.86.
\citet[][]{Scarlata2024} also challenged the Case B assumption using a single
galaxy at $z=0.0695$, for which they measured the ${\rm H\alpha}$/${\rm H\beta}$
ratio of 2.62 based on the $R\sim 867$ spectrum taken at the MMT. By considering
other line ratios, they suggested that the deviation is likely driven by 
physical mechanisms, most likely the Balmer self-absorption and scattering 
effects in non-spherically symmetric gas geometries. 
{In the context of non-Case B recombination, \citet[][]{Yanagisawa2024} 
examined two galaxies at $z\approx 6$ previously reported to have 
${\rm H\alpha}$/${\rm H\beta}$ line ratios less than 2.86 
\citep[][see also Appendix~\ref{appendix:literature}]{Cameron2024,Topping2024}. 
They proposed two possibilities for
explanation, namely, density-bounded nebulae that are opaque only up to around 
Ly$\gamma$--Ly8 transitions and ionization-bounded nebulae partly or fully 
surrounded by optically-thick, excited H I clouds. 
Lastly, \citet[][]{McClymont2025} used the public JADES spectroscopic catalog
(the same as the one that we used for comparison in Section 3.4) and found 
52 galaxies that have non-Case B Balmer decrements. They elaborated the scenarios 
of density-bounded and ionization-bounded nebulae and suggested that the former 
would provide a more natural explanation.
}

    Our result enhances the argument that the Case B recombination is not 
universally applicable. As compared to the study of \citet[][]{Pirzkal2024},
our result is drawn from a much enlarged sample (by $>4$$\times$) over a much 
wider redshift range ($1.5\leq z\leq 7$), and our line measurement, thanks to
the NIRSpec MSA mode, is not prone to the data reduction complications that
slitless spectroscopy usually suffers. 
{As compared to that of \citet[][]{McClymont2025}, we utilize both the JADES
and the CEERS data and have a larger sample. In addition, we also show that the
non-Case B Balmer decrements are also present in other large spectroscopic
surveys from the ground.} All in all, there is now sufficient
evidence that the Case B assumption does not universally hold.

    We note that the failure of the Case B assumption is not necessarily only
limited to the galaxies that have ${\rm H\alpha}$/${\rm H\beta}$ ratios less 
than 2.86, which is also a point raised in \citet[][]{Pirzkal2024} and 
\citet[][]{Scarlata2024}. For example, galaxies with intrinsic non-Case B 
Balmer decrements could have dust reddening that easily creates line ratios 
that do not seemingly violate the Case B limit and therefore are disguised. Our
community will need to reconsider how to assess dust reddening/extinction in 
extragalactic studies in general.

\subsection{Mysterious ${\rm H\alpha}$--[O~\Romannum{3}] Correlation}

    The mysteriously tight correlation between ${\rm H\alpha}$ and
[O~\Romannum{3}] is what we observe \emph{before} applying any dust extinction
correction, and therefore is independent of the non-Case B recombination 
problem discussed above. Based on the SED fitting, most of these objects have 
non-negligible dust reddening, and therefore the linearity observed can hardly
be explained. As shown in Figure~\ref{fig:line_relations}, the 
relation remains tight in four successive redshift bins, and there is no
obvious redshift evolution from $z\approx 2$ to $z\approx 7$.
In contrast, the results from the three ground-based redshift surveys that
we examined do not present the same picture: there is still a correlation, but
it has a much wider dispersion, as one would expect when considering dust
reddening. 

    Nevertheless, there is still room to reconcile these observations. The
vast majority of the galaxies from the three ground-based surveys have line 
fluxes brighter than $3\times10^{-17}$~erg~s$^{-1}$~cm$^{-2}$, while most of
the objects in the NIRSpec samples are fainter than this limit. Therefore, we 
surmise that there could be a transition for this relation at around this limit. 
Future NIRSpec data, when cumulating a sufficient number of
objects at brighter levels, will show whether the current discrepancy
with the ground-based results can be explained.

\section{Conclusion}\label{sec:conclusion}

    Using a large sample consisting of 251 $R\sim 100$ PRISM spectra and 312 
$R\sim 1000$ Grating spectra (totaling 480 unique galaxies) taken by the JWST 
NIRSpec instrument in three widely separated extragalactic fields,
we found a surprisingly tight linear relationship between ${\rm H\alpha}$ 
and [O~\Romannum{3}] emission line fluxes (in logarithmic space) that 
spans a wide redshift range ($z\sim 1.5-7$). Such a tight linear correlation is
mysterious, as dust reddening would have skewed any intrinsic correlation 
between the two. From a subsample consisting of 179 PRISM spectra and 265 
Grating spectra (378 unique galaxies) that allows the derivation of dust 
reddening using the ``gold standard'' Balmer decrement method, we found yet 
another surprising result: about {$\sim$30\%} of our objects have 
${\rm H\alpha}$/${\rm H\beta}$ line flux ratios smaller than the canonical 
Case~B value of 2.86, some of which are as small as $\sim$1.0. This casts 
serious doubts to the universality of the Case B assumption. We show that such 
non-Case~B ${\rm H\alpha}$/${\rm H\beta}$ ratios are also present in large 
fractions ($\sim$30\%) in at least two public JWST NIRSpec catalogs produced by 
two independent teams as well as in three ground-based redshift surveys carried 
out at Keck and Subaru. Our finding adds more weight to the recent challenges to 
the validity of the Case B assumption
\citep[][]{Pirzkal2024, Scarlata2024, McClymont2025}, and it is highly likely
that the Case B recombination does not hold universally. This will have a 
far-reaching impact to the community at large when carrying out spectral analysis 
in general.
{Encouragingly, some theoretical discussions have started to explain the
non-Case B Balmer decrements 
\citep[][]{Scarlata2024, Yanagisawa2024, McClymont2025}. However, much still needs
to be done to reach a consensus. In addition, the tight 
${\rm H\alpha}$--[O~\Romannum{3}] correlation also calls for theoretical studies 
to understand the underlying physical mechanism(s).}

\begin{acknowledgments}

   {The authors would like to thank the anonymous referee for the 
constructive comments.} 
We acknowledge the support from the NSF grant AST-2307447, the NASA 
grant 80NSSC23K0491 and the University of Missouri Research Council grant 
URC-23-029. This project is based on observations made with the NASA/ESA/CSA 
James Webb Space Telescope and obtained from the Mikulski Archive for Space 
Telescopes, which is a collaboration between the Space Telescope Science 
Institute (STScI/NASA), the Space Telescope European Coordinating Facility
(ST-ECF/ESA), and the Canadian Astronomy Data Centre (CADC/NRC/CSA). 
All the JWST NIRSpec data used in this paper can be found in MAST: 
\dataset[10.17909/mzqy-r528]{http://dx.doi.org/10.17909/mzqy-r528}.

\end{acknowledgments}

\appendix
\counterwithin{figure}{section}
\counterwithin{table}{section}

\section{Monte-Carlo Test for Statistical Bias in Balmer Decrement} 
\label{appendix:monte-carlo}

\begin{figure}[hbt!]
    \centering
    \includegraphics[width=0.49\linewidth]{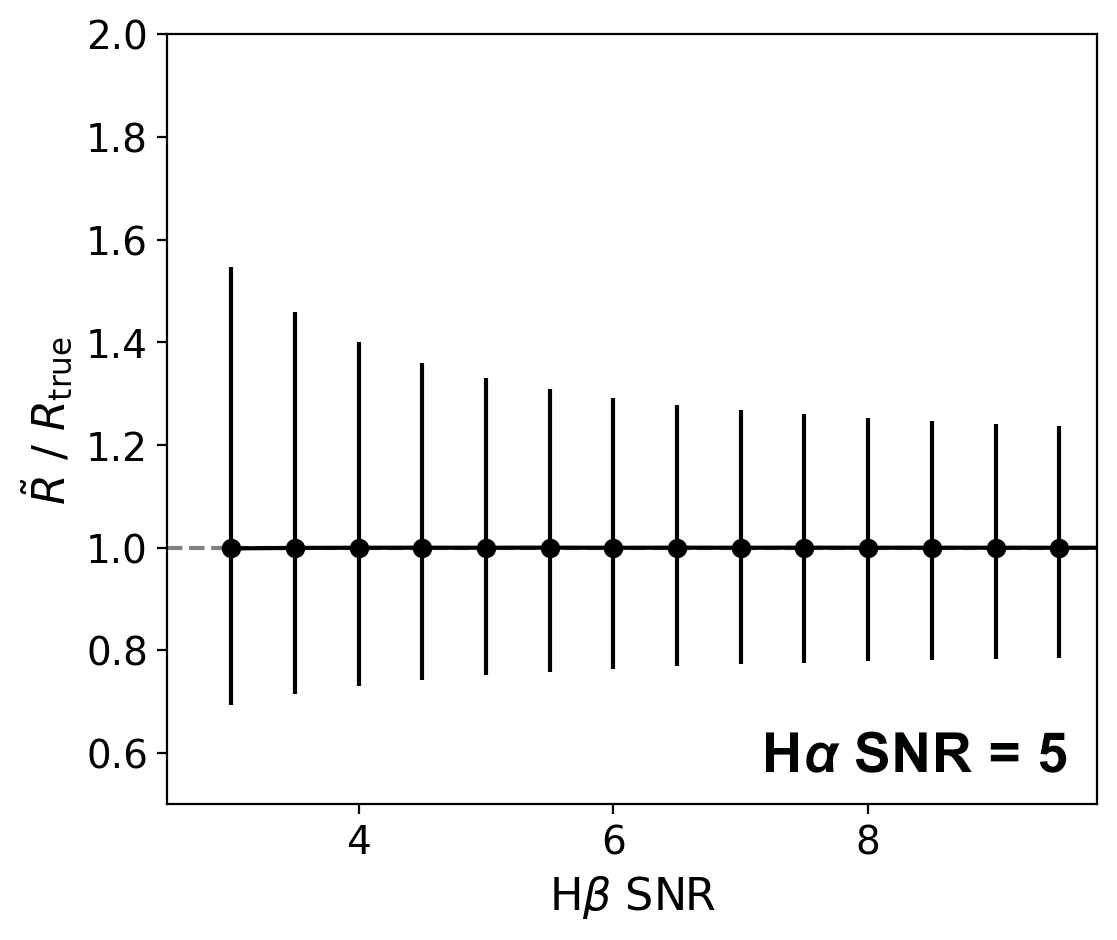}
    \includegraphics[width=0.49\linewidth]{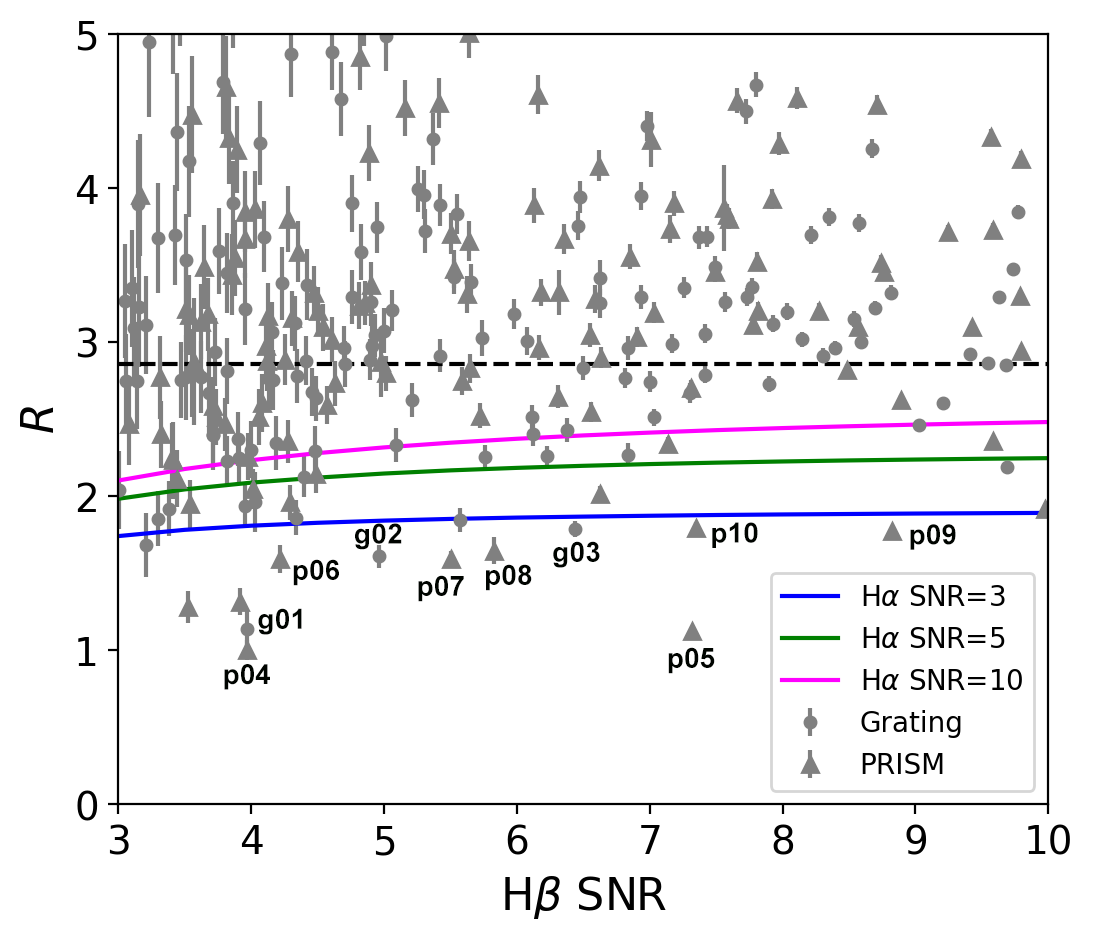}
    \caption{Left: Comparison of the median $\rm H\alpha/H\beta$ ratio
     $\tilde R$ to the fixed true value $R_{\rm true}=2.86$ as a function of 
     $\rm H\beta$ SNR, which are extracted from the MC test at a fixed 
     $\rm H\alpha$ SNR of 5. The error bars show the 16th to 84th percentile 
     range of the synthetic ratio distribution at each $\rm H\beta$ SNR.   
     Right: Balmer decrements of the objects in our sample (circles for the
     Grating set and triangles for the PRISM set) versus the measured 
     $\rm H\beta$ SNR. The dashed horizontal line indicates the canonical value 
     of 2.86. The blue, green and purple curves represent the 16th percentile 
     lower limit as derived from the MC test at the $\rm H\alpha$ SNR of 3, 5 
     and 10, respectively. Ten data points below the blue curve and at
     $\rm H\beta$ ${\rm SNR}>4$ are labeled, whose details are given in Table~\ref{tab:10dev}.
    }
    \label{fig:mc-test}
\end{figure}

    A concern in the statistics of ratios is that a ratio could be skewed when 
the denominator is the smaller quantity and has large errors. To quantify how 
this effect impacts our $\rm H\alpha/H\beta$ line ratio measurements, we 
carried out a Monte-Carlo (MC) test by fixing the $\rm H\alpha$ SNR 
at a series of certain values $G$ and varying 
the $\rm H\beta$ SNR. First, we generated $10^6$ realizations of $\rm H\alpha$ 
fluxes from a normal distribution with $\mu=2.86$ and $\sigma=2.86/G$ to fix 
their SNR at $G$ for the nominal line ratio of 
$R={\rm H\alpha/H\beta}=2.86$. 
Then, for each arbitrary $\rm H\beta$ SNR, we drew $10^6$ $\rm H\beta$ fluxes 
also from a normal distribution, but with $\mu=1$ and 
$\sigma=1/{\rm SNR_{H\beta}}$. From these drawings, we computed the synthetic 
$R={\rm H\alpha/H\beta}$ by randomly pairing each $\rm H\alpha$ flux 1,000 
times with a random $\rm H\beta$ flux in the drawing. We took the median of 
these ratios as the measured value $\tilde{R}$ that would be theoretically 
obtained in repeated observations, as well as the 16th and 84th percentiles to 
represent the lower and upper bounds. We then compared it to 
$R_{\rm true}=2.86$ by calculating $\tilde{R}/R_{\rm true}$. 

  The left panel of Figure~\ref{fig:mc-test} shows the resulting 
$\tilde{R}/R_{\rm true}$ as a function of $\rm H\beta$ SNR 
for the run at $G=5$ ($\rm H\alpha$ SNR of 5). 
This provides two insights. Firstly, a Balmer decrement higher than the canonical
value is more easily explained by the noise fluctuation than one that is lower. 
Secondly, even in the most extreme case at $\rm H\beta$ ${\rm SNR = 3}$), 
the 16th percentile of the synthetic $R$ distribution is 
$\sim0.7\times R_{\rm true}$, i.e., $0.7\times2.86\approx2.0$. 
In this work, however, we observed a large number of objects with measured 
Balmer decrements below this 16th percentile lower bound. To further
demonstrate this point, the right panel of Figure~\ref{fig:mc-test} shows
the Balmer decrements ($R$) of the objects in our sample versus the measured
$\rm H\beta$ SNR. The blue, green and purple curves represent the 16th 
percentile lower limit of $R$ as derived from our MC test at the $\rm H\alpha$ 
SNR of 3, 5 and 10, respectively. The blue curve is the worst-case scenario, as 
the lowest SNR of our $\rm H\alpha$ lines must be 3. Clearly, there are still a
significant number of objects falling below the blue curve. In fact, most of
them have $\rm H\alpha$ SNR much larger than 3, and their corresponding 16th
percentile low bounds of R are well above the blue curve. As examples, we
select 10 data points below the blue curve that have $\rm H\beta$
${\rm SNR} > 4$ (labeled in the figure) and provide the details in 
Table~\ref{tab:10dev}. These very low $R$ values can hardly be explained by the 
uncertainties in the line strength measurements.

   In summary, in case of a low H$\beta$ SNR, the statistical bias due to
measurement uncertainty alone cannot account for those low-ratio galaxies.

\begin{table}[hbt!]
    \centering
    \tiny
    \begin{tabular}{l|ccc|ccccccc} \hline 
        Source ID & g01 & g02 & g03 & p04 & p05 & p06 & p07 & p08 & p09 & p10 \\ \hline
        $\rm H\beta$ SNR & 4.0 & 5.0 & 6.4 & 4.0 & 7.3 & 4.2 & 5.5 & 5.8 & 8.8 & 7.4 \\ \hline 
        $\rm H\alpha$ SNR & 4.7 & 15.8 & 14.9 & 39.8 & 37.2 & 22.5 & 45.4 & 6.5 & 47.8 & 33.2 \\ \hline 
        $R$ & $1.13\pm0.12$ & $1.61\pm0.07$ & $1.79\pm0.05$ & $1.00\pm0.06$ & $1.12\pm0.02$ & $1.59\pm0.06$ & $1.59\pm0.03$ & $1.65\pm0.05$ & $1.77\pm0.01$ & $1.80\pm0.02$ \\ \hline 
        $R_{\rm low,16}$ & 2.07 & 2.36 & 2.26 & 2.29 & 2.51 & 2.30 & 2.42 & 2.27 & 2.56 & 2.51 \\ \hline 
    \end{tabular}
    \caption{Details of the ten deviant data points labeled in 
    Figure~\ref{tab:10dev}.
    $R$ is the measured Balmer decrement, and $R_{\rm low, 16}$ is the expected
    16th percentile lower bound (based on our MC simulation) corresponding to the given H$\beta$ and H$\alpha$ SNR.
    }
    \label{tab:10dev}
\end{table}

\section{Reported non-Case B Incidents in Literature}\label{appendix:literature}

   Several incidents of non-Case B Balmer decrements have been specifically 
reported in the literature, but {very few} of these were used to question the 
validity of the Case B assumption.

    While studying 24 ${\rm Ly\alpha}$ emitters at
$z\sim 0.2$--0.3, \citet[][]{Atek2009} found that a few of them have ``negative
extinction with large uncertainties'' and believed that these could be due to
calibration errors, a stronger stellar absorption than assumed, or enhanced
${\rm H\beta}$ emission from a reflection nebula. 
\citet[][]{Reddy2015} measured the Balmer decrements from the early observations 
of the MOSDEF program and showed that there were a significant number of sources
with non-Case B ratios, but this was not further discussed. In their study of 
``blueberry galaxies'' selected from the SDSS, \citet[][]{Yang2017} mentioned 
that most of them (32 out of 41) have non-Case B ratios; however, they attributed 
this to the poor flat-field calibration at the wavelength range where the 
${\rm H\alpha}$ line locates. 
{\citet[][]{Prescott2022} identified one galaxy (out of the total of 11 
studied) having an abnormal ${\rm H\alpha}$/${\rm H\beta}$ ratio of $\sim$2.0 
but without further questioning of the validity of Case B.}
Such sporadic discoveries continued into the JWST era. Using the 
wide-field slitless spectroscopy done by the NIRISS instrument, 
\citet[][]{Matharu2023} investigated spatially resolved Balmer decrements in 117 
galaxies by stacking and saw some indications of ${\rm H\alpha}$ and 
${\rm H\beta}$ line ratio being lower than 2.86 in the
inner parts of the galaxies. However, they concluded that such ratios were due
to spurious measurements. Two incidents of non-Case B line ratios
from the JWST NIRSpec medium-resolution grating data have also been reported. 
\citet[][]{Topping2024} studied a metal-poor $z=6.1$ galaxy, which has the 
line ratio of 2.55; they argued that this ratio could still be consistent with
the Case B recombination under twice higher electron temperature and two orders
of magnitude higher electron density. \citet[][]{Cameron2024} discussed in detail
a $z=5.943$ Ly$\alpha$ emitter that has a prominent nebular continuum (this 
galaxy is in our sample), for which their measured line fluxes give the 
${\rm H\alpha}$/${\rm H\beta}$ ratio of $\sim$2.65. However, they believed
that this was due to the underestimated ${\rm H\alpha}$ flux in their 
measurement. 
{\citet[][]{Sandles2024} conducted a dedicated study of Balmer decrements 
using the NIRSpec PRISM spectra in the JADES fields. Their sample includes about 
a dozen galaxies with ${\rm H\alpha}$/${\rm H\beta}$ ratios less than 2.86 (see 
their Figure 2), however these were not discussed as violations of Case B
recombination.}


\bibliography{sample631}{}

\begin{thebibliography}{}
\expandafter\ifx\csname natexlab\endcsname\relax\def\natexlab#1{#1}\fi
\providecommand{\url}[1]{\href{#1}{#1}}
\providecommand{\dodoi}[1]{doi:~\href{http://doi.org/#1}{\nolinkurl{#1}}}
\providecommand{\doeprint}[1]{\href{http://ascl.net/#1}{\nolinkurl{http://ascl.net/#1}}}
\providecommand{\doarXiv}[1]{\href{https://arxiv.org/abs/#1}{\nolinkurl{https://arxiv.org/abs/#1}}}

\bibitem[{P. {Arrabal Haro} {et~al.}(2023){Arrabal Haro}, {Dickinson},
  {Finkelstein}, {Fujimoto}, {Fern{\'a}ndez}, {Kartaltepe}, {Jung}, {Cole},
  {Burgarella}, {Chworowsky}, {Hutchison}, {Morales}, {Papovich}, {Simons},
  {Amor{\'\i}n}, {Backhaus}, {Bagley}, {Bisigello}, {Calabr{\`o}},
  {Castellano}, {Cleri}, {Dav{\'e}}, {Dekel}, {Ferguson}, {Fontana}, {Gawiser},
  {Giavalisco}, {Harish}, {Hathi}, {Hirschmann}, {Holwerda}, {Huertas-Company},
  {Koekemoer}, {Larson}, {Lucas}, {Mobasher}, {P{\'e}rez-Gonz{\'a}lez},
  {Pirzkal}, {Rose}, {Santini}, {Trump}, {de la Vega}, {Wang}, {Weiner},
  {Wilkins}, {Yang}, {Yung}, \& {Zavala}}]{Haro2023}
{Arrabal Haro}, P., {Dickinson}, M., {Finkelstein}, S.~L., {et~al.} 2023,
  \bibinfo{title}{{Spectroscopic Confirmation of CEERS NIRCam-selected Galaxies
  at z ≃ 8-10},} \apjl, 951, L22, \dodoi{10.3847/2041-8213/acdd54}

\bibitem[{H. {Atek} {et~al.}(2009){Atek}, {Kunth}, {Schaerer}, {Hayes},
  {Deharveng}, {{\"O}stlin}, \& {Mas-Hesse}}]{Atek2009}
{Atek}, H., {Kunth}, D., {Schaerer}, D., {et~al.} 2009,
  \bibinfo{title}{{Empirical estimate of Ly{\ensuremath{\alpha}} escape
  fraction in a statistical sample of Ly{\ensuremath{\alpha}} emitters},} \aap,
  506, L1, \dodoi{10.1051/0004-6361/200912787}

\bibitem[{J.~A. {Baldwin} {et~al.}(1981){Baldwin}, {Phillips}, \&
  {Terlevich}}]{Baldwin1981}
{Baldwin}, J.~A., {Phillips}, M.~M., \& {Terlevich}, R. 1981,
  \bibinfo{title}{{Classification parameters for the emission-line spectra of
  extragalactic objects.},} \pasp, 93, 5, \dodoi{10.1086/130766}

\bibitem[{R. {Bezanson} {et~al.}(2024){Bezanson}, {Labbe}, {Whitaker}, {Leja},
  {Price}, {Franx}, {Brammer}, {Marchesini}, {Zitrin}, {Wang}, {Weaver},
  {Furtak}, {Atek}, {Coe}, {Cutler}, {Dayal}, {van Dokkum}, {Feldmann},
  {F{\"o}rster Schreiber}, {Fujimoto}, {Geha}, {Glazebrook}, {de Graaff},
  {Greene}, {Juneau}, {Kassin}, {Kriek}, {Khullar}, {Maseda}, {Mowla},
  {Muzzin}, {Nanayakkara}, {Nelson}, {Oesch}, {Pacifici}, {Pan}, {Papovich},
  {Setton}, {Shapley}, {Smit}, {Stefanon}, {Taylor}, \&
  {Williams}}]{Bezanson2024_uncover}
{Bezanson}, R., {Labbe}, I., {Whitaker}, K.~E., {et~al.} 2024,
  \bibinfo{title}{{The JWST UNCOVER Treasury Survey: Ultradeep NIRSpec and
  NIRCam Observations before the Epoch of Reionization},} \apj, 974, 92,
  \dodoi{10.3847/1538-4357/ad66cf}

\bibitem[{G. {Brammer}(2023){Brammer}}]{Brammar23_msaexp}
{Brammer}, G. 2023, \bibinfo{title}{{msaexp: NIRSpec analyis tools},}, 0.6.17
  Zenodo, \dodoi{10.5281/zenodo.7299500}

\bibitem[{G. {Bruzual} \& S. {Charlot}(2003){Bruzual} \&
  {Charlot}}]{Bruzual2003}
{Bruzual}, G., \& {Charlot}, S. 2003, \bibinfo{title}{{Stellar population
  synthesis at the resolution of 2003},} \mnras, 344, 1000,
  \dodoi{10.1046/j.1365-8711.2003.06897.x}

\bibitem[{H. {Bushouse} {et~al.}(2024){Bushouse}, {Eisenhamer}, {Dencheva},
  {Davies}, {Greenfield}, {Morrison}, {Hodge}, {Simon}, {Grumm}, {Droettboom},
  {Slavich}, {Sosey}, {Pauly}, {Miller}, {Jedrzejewski}, {Hack}, {Davis},
  {Crawford}, {Law}, {Gordon}, {Regan}, {Cara}, {MacDonald}, {Bradley},
  {Shanahan}, {Jamieson}, {Teodoro}, {Williams}, \&
  {Pena-Guerrero}}]{Bushouse24_jwppl}
{Bushouse}, H., {Eisenhamer}, J., {Dencheva}, N., {et~al.} 2024,
  \bibinfo{title}{{JWST Calibration Pipeline},}, 1.14.0 Zenodo,
  \dodoi{10.5281/zenodo.6984365}

\bibitem[{D. {Calzetti} {et~al.}(2000){Calzetti}, {Armus}, {Bohlin}, {Kinney},
  {Koornneef}, \& {Storchi-Bergmann}}]{Calzetti2000}
{Calzetti}, D., {Armus}, L., {Bohlin}, R.~C., {et~al.} 2000,
  \bibinfo{title}{{The Dust Content and Opacity of Actively Star-forming
  Galaxies},} \apj, 533, 682, \dodoi{10.1086/308692}

\bibitem[{D. {Calzetti} {et~al.}(1994){Calzetti}, {Kinney}, \&
  {Storchi-Bergmann}}]{Calzetti1994}
{Calzetti}, D., {Kinney}, A.~L., \& {Storchi-Bergmann}, T. 1994,
  \bibinfo{title}{{Dust Extinction of the Stellar Continua in Starburst
  Galaxies: The Ultraviolet and Optical Extinction Law},} \apj, 429, 582,
  \dodoi{10.1086/174346}

\bibitem[{A.~J. {Cameron} {et~al.}(2024){Cameron}, {Katz}, {Witten}, {Saxena},
  {Laporte}, \& {Bunker}}]{Cameron2024}
{Cameron}, A.~J., {Katz}, H., {Witten}, C., {et~al.} 2024,
  \bibinfo{title}{{Nebular dominated galaxies: insights into the stellar
  initial mass function at high redshift},} \mnras, 534, 523,
  \dodoi{10.1093/mnras/stae1547}

\bibitem[{A.~C. {Carnall} {et~al.}(2018){Carnall}, {McLure}, {Dunlop}, \&
  {Dav{\'e}}}]{Carnall18_bagpipes}
{Carnall}, A.~C., {McLure}, R.~J., {Dunlop}, J.~S., \& {Dav{\'e}}, R. 2018,
  \bibinfo{title}{{Inferring the star formation histories of massive quiescent
  galaxies with BAGPIPES: evidence for multiple quenching mechanisms},} \mnras,
  480, 4379, \dodoi{10.1093/mnras/sty2169}

\bibitem[{F. {D'Eugenio} {et~al.}(2024){D'Eugenio}, {Cameron}, {Scholtz},
  {Carniani}, {Willott}, {Curtis-Lake}, {Bunker}, {Parlanti}, {Maiolino},
  {Willmer}, {Jakobsen}, {Robertson}, {Johnson}, {Tacchella}, {Cargile},
  {Rawle}, {Arribas}, {Chevallard}, {Curti}, {Egami}, {Eisenstein}, {Kumari},
  {Looser}, {Rieke}, {Rodr{\'\i}guez Del Pino}, {Saxena}, {{\"U}bler},
  {Venturi}, {Witstok}, {Baker}, {Bhatawdekar}, {Bonaventura}, {Boyett},
  {Charlot}, {Danhaive}, {Hainline}, {Hausen}, {Helton}, {Ji}, {Ji}, {Jones},
  {Joud{\v{z}}balis}, {Maseda}, {P{\'e}rez-Gonz{\'a}lez}, {Perna},
  {Pusk{\'a}s}, {Shivaei}, {Silcock}, {Simmonds}, {Smit}, {Sun}, {Villanueva},
  {Williams}, \& {Zhu}}]{DEugenio24_jadesdr3}
{D'Eugenio}, F., {Cameron}, A.~J., {Scholtz}, J., {et~al.} 2024,
  \bibinfo{title}{{JADES Data Release 3 -- NIRSpec/MSA spectroscopy for 4,000
  galaxies in the GOODS fields},} arXiv e-prints, arXiv:2404.06531,
  \dodoi{10.48550/arXiv.2404.06531}

\bibitem[{N. Earl {et~al.}(2024)Earl, Tollerud, O'Steen, brechmos, Kerzendorf,
  Busko, shaileshahuja, D'Avella, Lim, Robitaille, Ginsburg, Homeier, Sipőcz,
  Averbukh, Tocknell, Cherinka, Ogaz, Geda, Conroy, Davies, Günther, Barbary,
  Foster, Droettboom, Nguyen, Bray, Casey, Cruz, Ferguson, \&
  Crawford}]{Earl24_specutils}
Earl, N., Tollerud, E., O'Steen, R., {et~al.} 2024,
  \bibinfo{title}{astropy/specutils: v1.19.0,}, v1.19.0 Zenodo,
  \dodoi{10.5281/zenodo.14042033}

\bibitem[{D.~J. {Eisenstein} {et~al.}(2023){Eisenstein}, {Johnson},
  {Robertson}, {Tacchella}, {Hainline}, {Jakobsen}, {Maiolino}, {Bonaventura},
  {Bunker}, {Cameron}, {Cargile}, {Curtis-Lake}, {Hausen}, {Pusk{\'a}s},
  {Rieke}, {Sun}, {Willmer}, {Willott}, {Alberts}, {Arribas}, {Baker}, {Baum},
  {Bhatawdekar}, {Carniani}, {Charlot}, {Chen}, {Chevallard}, {Curti},
  {DeCoursey}, {D'Eugenio}, {de Graaff}, {Egami}, {Helton}, {Ji}, {Jones},
  {Kumari}, {L{\"u}tzgendorf}, {Laseter}, {Looser}, {Lyu}, {Maseda}, {Nelson},
  {Parlanti}, {Rauscher}, {Rawle}, {Rieke}, {Rix}, {Rujopakarn}, {Sandles},
  {Saxena}, {Scholtz}, {Sharpe}, {Shivaei}, {Simmonds}, {Smit}, {Topping},
  {{\"U}bler}, {Venturi}, {Williams}, {Witstok}, \& {Woodrum}}]{Eisenstein2023}
{Eisenstein}, D.~J., {Johnson}, B.~D., {Robertson}, B., {et~al.} 2023,
  \bibinfo{title}{{The JADES Origins Field: A New JWST Deep Field in the JADES
  Second NIRCam Data Release},} arXiv e-prints, arXiv:2310.12340,
  \dodoi{10.48550/arXiv.2310.12340}

\bibitem[{M. {Figueira} {et~al.}(2022){Figueira}, {Pollo}, {Ma{\l}ek}, {Buat},
  {Boquien}, {Pistis}, {Cassar{\`a}}, {Vergani}, {Hamed}, \&
  {Salim}}]{Figueira2022}
{Figueira}, M., {Pollo}, A., {Ma{\l}ek}, K., {et~al.} 2022,
  \bibinfo{title}{{SFR estimations from z = 0 to z = 0.9. A comparison of SFR
  calibrators for star-forming galaxies},} \aap, 667, A29,
  \dodoi{10.1051/0004-6361/202141701}

\bibitem[{S.~L. {Finkelstein} {et~al.}(2025){Finkelstein}, {Bagley}, {Arrabal
  Haro}, {Dickinson}, {Ferguson}, {Kartaltepe}, {Kocevski}, {Koekemoer},
  {Lotz}, {Papovich}, {Perez-Gonzalez}, {Pirzkal}, {Somerville}, {Trump},
  {Yang}, {Yung}, {Fontana}, {Grazian}, {Grogin}, {Kewley}, {Kirkpatrick},
  {Larson}, {Pentericci}, {Ravindranath}, {Wilkins}, {Almaini}, {Amorin},
  {Barro}, {Bhatawdekar}, {Bisigello}, {Brooks}, {Buitrago}, {Calabro},
  {Castellano}, {Cheng}, {Cleri}, {Cole}, {Cooper}, {Cooper}, {Costantin},
  {Cox}, {Croton}, {Daddi}, {Davis}, {Dekel}, {Elbaz}, {Fernandez}, {Fujimoto},
  {Gandolfi}, {Gardner}, {Gawiser}, {Giavalisco}, {Gomez-Guijarro}, {Guo},
  {Gupta}, {Hathi}, {Harish}, {Henry}, {Hirschmann}, {Hu}, {Hutchison}, {Iyer},
  {Jaskot}, {Jha}, {Jung}, {Kokorev}, {Kurczynski}, {Leung}, {Llerena}, {Long},
  {Lucas}, {Lu}, {McGrath}, {McIntosh}, {Merlin}, {Morales}, {Napolitano},
  {Pacucci}, {Pandya}, {Rafelski}, {Rodighiero}, {Rose}, {Santini}, {Seille},
  {Simons}, {Shen}, {Straughn}, {Tacchella}, {Vanderhoof}, {Vega-Ferrero},
  {Weiner}, {Willmer}, {Zhu}, {Bell}, {Wuyts}, {Holwerda}, {Wang}, {Wang}, \&
  {Zavala}}]{Finkelstein2025}
{Finkelstein}, S.~L., {Bagley}, M.~B., {Arrabal Haro}, P., {et~al.} 2025,
  \bibinfo{title}{{The Cosmic Evolution Early Release Science Survey (CEERS)},}
  arXiv e-prints, arXiv:2501.04085, \dodoi{10.48550/arXiv.2501.04085}

\bibitem[{D. {Kashino} {et~al.}(2019){Kashino}, {Silverman}, {Sanders},
  {Kartaltepe}, {Daddi}, {Renzini}, {Rodighiero}, {Puglisi}, {Valentino},
  {Juneau}, {Arimoto}, {Nagao}, {Ilbert}, {Le F{\`e}vre}, \&
  {Koekemoer}}]{Kashino2019}
{Kashino}, D., {Silverman}, J.~D., {Sanders}, D., {et~al.} 2019,
  \bibinfo{title}{{The FMOS-COSMOS Survey of Star-forming Galaxies at z
  {\ensuremath{\sim}} 1.6. VI. Redshift and Emission-line Catalog and Basic
  Properties of Star-forming Galaxies},} \apjs, 241, 10,
  \dodoi{10.3847/1538-4365/ab06c4}

\bibitem[{R.~C. {Kennicutt}(1998){Kennicutt}}]{Kennicutt1998}
{Kennicutt}, Jr., R.~C. 1998, \bibinfo{title}{{Star Formation in Galaxies Along
  the Hubble Sequence},} \araa, 36, 189, \dodoi{10.1146/annurev.astro.36.1.189}

\bibitem[{L.~J. {Kewley} \& M.~A. {Dopita}(2002){Kewley} \& {Dopita}}]{KD02}
{Kewley}, L.~J., \& {Dopita}, M.~A. 2002, \bibinfo{title}{{Using Strong Lines
  to Estimate Abundances in Extragalactic H II Regions and Starburst
  Galaxies},} \apjs, 142, 35, \dodoi{10.1086/341326}

\bibitem[{L.~J. {Kewley} {et~al.}(2019){Kewley}, {Nicholls}, \&
  {Sutherland}}]{Kewley2019}
{Kewley}, L.~J., {Nicholls}, D.~C., \& {Sutherland}, R.~S. 2019,
  \bibinfo{title}{{Understanding Galaxy Evolution Through Emission Lines},}
  \araa, 57, 511, \dodoi{10.1146/annurev-astro-081817-051832}

\bibitem[{M. {Kriek} {et~al.}(2015){Kriek}, {Shapley}, {Reddy}, {Siana},
  {Coil}, {Mobasher}, {Freeman}, {de Groot}, {Price}, {Sanders}, {Shivaei},
  {Brammer}, {Momcheva}, {Skelton}, {van Dokkum}, {Whitaker}, {Aird}, {Azadi},
  {Kassis}, {Bullock}, {Conroy}, {Dav{\'e}}, {Kere{\v{s}}}, \&
  {Krumholz}}]{Kriek15_mosdef}
{Kriek}, M., {Shapley}, A.~E., {Reddy}, N.~A., {et~al.} 2015,
  \bibinfo{title}{{The MOSFIRE Deep Evolution Field (MOSDEF) Survey: Rest-frame
  Optical Spectroscopy for \raisebox{-0.5ex}\textasciitilde1500 H-selected
  Galaxies at 1.37 < z < 3.8},} \apjs, 218, 15,
  \dodoi{10.1088/0067-0049/218/2/15}

\bibitem[{P. {Kroupa}(2001){Kroupa}}]{Kroupa2001}
{Kroupa}, P. 2001, \bibinfo{title}{{On the variation of the initial mass
  function},} \mnras, 322, 231, \dodoi{10.1046/j.1365-8711.2001.04022.x}

\bibitem[{R. {Maiolino} {et~al.}(2008){Maiolino}, {Nagao}, {Grazian},
  {Cocchia}, {Marconi}, {Mannucci}, {Cimatti}, {Pipino}, {Ballero}, {Calura},
  {Chiappini}, {Fontana}, {Granato}, {Matteucci}, {Pastorini}, {Pentericci},
  {Risaliti}, {Salvati}, \& {Silva}}]{Maiolino2008}
{Maiolino}, R., {Nagao}, T., {Grazian}, A., {et~al.} 2008,
  \bibinfo{title}{{AMAZE. I. The evolution of the mass-metallicity relation at
  z > 3},} \aap, 488, 463, \dodoi{10.1051/0004-6361:200809678}

\bibitem[{J. {Matharu} {et~al.}(2023){Matharu}, {Muzzin}, {Sarrouh}, {Brammer},
  {Abraham}, {Asada}, {Brada{\v{c}}}, {Desprez}, {Martis}, {Mowla}, {Noirot},
  {Sawicki}, {Strait}, {Willott}, {Gould}, {Grindlay}, \&
  {Harshan}}]{Matharu2023}
{Matharu}, J., {Muzzin}, A., {Sarrouh}, G. T.~E., {et~al.} 2023,
  \bibinfo{title}{{A First Look at Spatially Resolved Balmer Decrements at 1.0
  < z < 2.4 from JWST NIRISS Slitless Spectroscopy},} \apjl, 949, L11,
  \dodoi{10.3847/2041-8213/acd1db}

\bibitem[{W. {McClymont} {et~al.}(2025){McClymont}, {Tacchella}, {D'Eugenio},
  {Witten}, {Ji}, {Smith}, {Maiolino}, {Arribas}, {Scholtz}, {Simmonds}, \&
  {Witstok}}]{McClymont2025}
{McClymont}, W., {Tacchella}, S., {D'Eugenio}, F., {et~al.} 2025,
  \bibinfo{title}{{The density-bounded twilight of starbursts in the early
  Universe},} \mnras, \dodoi{10.1093/mnras/staf745}

\bibitem[{J. {Moustakas} {et~al.}(2006){Moustakas}, {Kennicutt}, \&
  {Tremonti}}]{Moustakas2006}
{Moustakas}, J., {Kennicutt}, Jr., R.~C., \& {Tremonti}, C.~A. 2006,
  \bibinfo{title}{{Optical Star Formation Rate Indicators},} \apj, 642, 775,
  \dodoi{10.1086/500964}

\bibitem[{T. {Nanayakkara} {et~al.}(2016){Nanayakkara}, {Glazebrook},
  {Kacprzak}, {Yuan}, {Tran}, {Spitler}, {Kewley}, {Straatman}, {Cowley},
  {Fisher}, {Labbe}, {Tomczak}, {Allen}, \& {Alcorn}}]{Nanayakkara2016}
{Nanayakkara}, T., {Glazebrook}, K., {Kacprzak}, G.~G., {et~al.} 2016,
  \bibinfo{title}{{ZFIRE: A KECK/MOSFIRE Spectroscopic Survey of Galaxies in
  Rich Environments at z {\ensuremath{\sim}} 2},} \apj, 828, 21,
  \dodoi{10.3847/0004-637X/828/1/21}

\bibitem[{D.~E. {Osterbrock}(1989){Osterbrock}}]{Osterbrock1989}
{Osterbrock}, D.~E. 1989, {Astrophysics of gaseous nebulae and active galactic
  nuclei}

\bibitem[{N. {Pirzkal} {et~al.}(2024){Pirzkal}, {Rothberg}, {Papovich}, {Shen},
  {Leung}, {Bagley}, {Finkelstein}, {Vanderhoof}, {Lotz}, {Koekemoer}, {Hathi},
  {Cheng}, {Cleri}, {Grogin}, {Yung}, {Dickinson}, {Ferguson}, {Gardner},
  {Jung}, {Kartaltepe}, {Ryan}, {Simons}, {Ravindranath}, {Berg}, {Backhaus},
  {Casey}, {Castellano}, {Ch{\'a}vez Ortiz}, {Chworowsky}, {Cox}, {Dav{\'e}},
  {Davis}, {Estrada-Carpenter}, {Fontana}, {Fujimoto}, {Giavalisco}, {Grazian},
  {Hutchison}, {Jaskot}, {Kewley}, {Kirkpatrick}, {Kocevski}, {Larson},
  {Matharu}, {Natarajan}, {Pentericci}, {P{\'e}rez-Gonz{\'a}lez}, {Snyder},
  {Somerville}, {Trump}, \& {Wilkins}}]{Pirzkal2024}
{Pirzkal}, N., {Rothberg}, B., {Papovich}, C., {et~al.} 2024,
  \bibinfo{title}{{The Next Generation Deep Extragalactic Exploratory Public
  Near-infrared Slitless Survey Epoch 1 (NGDEEP-NISS1): Extragalactic
  Star-formation and Active Galactic Nuclei at 0.5 < z < 3.6},} \apj, 969, 90,
  \dodoi{10.3847/1538-4357/ad429c}

\bibitem[{M.~K.~M. {Prescott} {et~al.}(2022){Prescott}, {Finlator}, {Cleri},
  {Trump}, \& {Papovich}}]{Prescott2022}
{Prescott}, M. K.~M., {Finlator}, K.~M., {Cleri}, N.~J., {Trump}, J.~R., \&
  {Papovich}, C. 2022, \bibinfo{title}{{Using Multiple Emission Line Ratios to
  Constrain the Slope of the Dust Attenuation Law},} \apj, 928, 71,
  \dodoi{10.3847/1538-4357/ac5459}

\bibitem[{S.~H. {Price} {et~al.}(2024){Price}, {Bezanson}, {Labbe}, {Furtak},
  {de Graaff}, {Greene}, {Kokorev}, {Setton}, {Suess}, {Brammer}, {Cutler},
  {Leja}, {Pan}, {Wang}, {Weaver}, {Whitaker}, {Atek}, {Burgasser},
  {Chemerynska}, {Dayal}, {Feldmann}, {F{\"o}rster Schreiber}, {Fudamoto},
  {Fujimoto}, {Glazebrook}, {Goulding}, {Khullar}, {Kriek}, {Marchesini},
  {Maseda}, {Miller}, {Muzzin}, {Nanayakkara}, {Nelson}, {Oesch}, {Shipley},
  {Smit}, {Taylor}, {van Dokkum}, {Williams}, \&
  {Zitrin}}]{Price2024_uncover-spec}
{Price}, S.~H., {Bezanson}, R., {Labbe}, I., {et~al.} 2024,
  \bibinfo{title}{{The UNCOVER Survey: First Release of Ultradeep JWST/NIRSpec
  PRISM spectra for \raisebox{-0.5ex}\textasciitilde700 galaxies from
  z\raisebox{-0.5ex}\textasciitilde0.3-13 in Abell 2744},} arXiv e-prints,
  arXiv:2408.03920, \dodoi{10.48550/arXiv.2408.03920}

\bibitem[{N.~A. {Reddy} {et~al.}(2015){Reddy}, {Kriek}, {Shapley}, {Freeman},
  {Siana}, {Coil}, {Mobasher}, {Price}, {Sanders}, \& {Shivaei}}]{Reddy2015}
{Reddy}, N.~A., {Kriek}, M., {Shapley}, A.~E., {et~al.} 2015,
  \bibinfo{title}{{The MOSDEF Survey: Measurements of Balmer Decrements and the
  Dust Attenuation Curve at Redshifts z \raisebox{-0.5ex}\textasciitilde
  1.4-2.6},} \apj, 806, 259, \dodoi{10.1088/0004-637X/806/2/259}

\bibitem[{L. {Sandles} {et~al.}(2024){Sandles}, {D'Eugenio}, {Maiolino},
  {Looser}, {Arribas}, {Baker}, {Bonaventura}, {Bunker}, {Cameron}, {Carniani},
  {Charlot}, {Chevallard}, {Curti}, {Curtis-Lake}, {de Graaff}, {Eisenstein},
  {Hainline}, {Ji}, {Johnson}, {Jones}, {Kumari}, {Nelson}, {Perna}, {Rawle},
  {Rix}, {Robertson}, {Del Pino}, {Scholtz}, {Shivaei}, {Smit}, {Sun},
  {Tacchella}, {{\"U}bler}, {Williams}, {Willott}, \& {Witstok}}]{Sandles2024}
{Sandles}, L., {D'Eugenio}, F., {Maiolino}, R., {et~al.} 2024,
  \bibinfo{title}{{JADES: Balmer decrement measurements at redshifts 4 < z <
  7},} \aap, 691, A305, \dodoi{10.1051/0004-6361/202347119}

\bibitem[{C. {Scarlata} {et~al.}(2024){Scarlata}, {Hayes}, {Panagia}, {Mehta},
  {Haardt}, \& {Bagley}}]{Scarlata2024}
{Scarlata}, C., {Hayes}, M., {Panagia}, N., {et~al.} 2024, \bibinfo{title}{{On
  the universal validity of Case B recombination theory},} arXiv e-prints,
  arXiv:2404.09015, \dodoi{10.48550/arXiv.2404.09015}

\bibitem[{B. {Sun} \& H. {Yan}(2025){Sun} \& {Yan}}]{SY2025a}
{Sun}, B., \& {Yan}, H. 2025, \bibinfo{title}{{On The Very Bright Dropouts
  Selected Using the James Webb Space Telescope NIRCam Instrument},} arXiv
  e-prints, arXiv:2502.05751, \dodoi{10.48550/arXiv.2502.05751}

\bibitem[{T.~L. {Suzuki} {et~al.}(2016){Suzuki}, {Kodama}, {Sobral},
  {Khostovan}, {Hayashi}, {Shimakawa}, {Koyama}, {Tadaki}, {Tanaka}, {Minowa},
  {Yamamoto}, {Smail}, \& {Best}}]{Suzuki2016}
{Suzuki}, T.~L., {Kodama}, T., {Sobral}, D., {et~al.} 2016, \bibinfo{title}{{[O
  III] emission line as a tracer of star-forming galaxies at high redshifts:
  comparison between H{\ensuremath{\alpha}} and [O III] emitters at z=2.23 in
  HiZELS},} \mnras, 462, 181, \dodoi{10.1093/mnras/stw1655}

\bibitem[{M.~W. {Topping} {et~al.}(2024){Topping}, {Stark}, {Senchyna}, {Plat},
  {Zitrin}, {Endsley}, {Charlot}, {Furtak}, {Maseda}, {Smit}, {Mainali},
  {Chevallard}, {Molyneux}, \& {Rigby}}]{Topping2024}
{Topping}, M.~W., {Stark}, D.~P., {Senchyna}, P., {et~al.} 2024,
  \bibinfo{title}{{Metal-poor star formation at z $>$ 6 with JWST: new insight
  into hard radiation fields and nitrogen enrichment on 20 pc scales},} \mnras,
  529, 3301, \dodoi{10.1093/mnras/stae682}

\bibitem[{J.~A. {Villa-V{\'e}lez} {et~al.}(2021){Villa-V{\'e}lez}, {Buat},
  {Theul{\'e}}, {Boquien}, \& {Burgarella}}]{Villavelez2021}
{Villa-V{\'e}lez}, J.~A., {Buat}, V., {Theul{\'e}}, P., {Boquien}, M., \&
  {Burgarella}, D. 2021, \bibinfo{title}{{Fitting spectral energy distributions
  of FMOS-COSMOS emission-line galaxies at z {\ensuremath{\sim}} 1.6: Star
  formation rates, dust attenuation, and [OIII]{\ensuremath{\lambda}}5007
  emission-line luminosities},} \aap, 654, A153,
  \dodoi{10.1051/0004-6361/202140890}

\bibitem[{R. {Wen} {et~al.}(2022){Wen}, {An}, {Zheng}, {Shi}, {Qin},
  {Gonzalez}, {Bian}, {Xu}, {Pan}, {Tan}, {Liu}, {Fang}, {Ren}, {Zhang},
  {Qiao}, \& {Liu}}]{Wen2022}
{Wen}, R., {An}, F., {Zheng}, X.~Z., {et~al.} 2022, \bibinfo{title}{{The
  Physical Properties of Star-forming Galaxies with Strong [O III] Lines at z =
  3.25},} \apj, 933, 50, \dodoi{10.3847/1538-4357/ac7392}

\bibitem[{H. {Yanagisawa} {et~al.}(2024){Yanagisawa}, {Ouchi}, {Nakajima},
  {Yajima}, {Umeda}, {Baba}, {Nakagawa}, {Nakane}, {Matsumoto}, {Ono},
  {Harikane}, {Isobe}, {Xu}, \& {Zhang}}]{Yanagisawa2024}
{Yanagisawa}, H., {Ouchi}, M., {Nakajima}, K., {et~al.} 2024,
  \bibinfo{title}{{Balmer Decrement Anomalies in Galaxies at z
  {\ensuremath{\sim}} 6 Found by JWST Observations: Density-bounded Nebulae or
  Excited H I Clouds?},} \apj, 974, 180, \dodoi{10.3847/1538-4357/ad7097}

\bibitem[{H. {Yang} {et~al.}(2017){Yang}, {Malhotra}, {Rhoads}, \&
  {Wang}}]{Yang2017}
{Yang}, H., {Malhotra}, S., {Rhoads}, J.~E., \& {Wang}, J. 2017,
  \bibinfo{title}{{Blueberry Galaxies: The Lowest Mass Young Starbursts},}
  \apj, 847, 38, \dodoi{10.3847/1538-4357/aa8809}

\end{thebibliography}
\bibliographystyle{aasjournal}

\end{document}